\documentclass{aastex7}

\usepackage{hyperref}
\usepackage{caption}

\begin{document}

\title{Differential Reddening and Extinction Law Analyses of Galactic Open Clusters}

\correspondingauthor{Shu Wang, Xiaodian Chen}
\email{shuwang@nao.cas.cn, chenxiaodian@nao.cas.cn}

\author{Xiaohan Chen}
\affiliation{CAS Key Laboratory of Optical Astronomy, National Astronomical Observatories, Chinese Academy of Sciences, Beijing 100101, China}
\affiliation{School of Astronomy and Space Science, University of the Chinese Academy of Sciences, Beijing, 100049, China}
\email{chenxh@bao.ac.cn}

\author[0000-0003-4489-9794]{Shu Wang}
\affiliation{CAS Key Laboratory of Optical Astronomy, National Astronomical Observatories, Chinese Academy of Sciences, Beijing 100101, China}
\affiliation{School of Astronomy and Space Science, University of the Chinese Academy of Sciences, Beijing, 100049, China}
\email{shuwang@nao.cas.cn}

\author[0000-0001-7084-0484]{Xiaodian Chen}
\affiliation{CAS Key Laboratory of Optical Astronomy, National Astronomical Observatories, Chinese Academy of Sciences, Beijing 100101, China}
\affiliation{School of Astronomy and Space Science, University of the Chinese Academy of Sciences, Beijing, 100049, China}
\affiliation{Institute for Frontiers in Astronomy and Astrophysics, Beijing Normal University, Beijing 102206, China}
\email{chenxiaodian@nao.cas.cn}

\begin{abstract}

Extinction significantly affects open cluster parameters and their use in studies of Galactic structure, yet homogeneous large sample measurements of open cluster extinction properties remain limited. Using \textit{Gaia}-era open cluster member samples combined with multi-band photometry and stellar parameters, we derive color excesses of member stars and provide the homogeneous characterization of the mean reddening, differential reddening, and color excess ratio (CER) at the cluster scale. 
Differential reddening increases systematically with mean reddening, with highly reddened clusters near the Galactic plane showing stronger extinction variations. 
Star-by-star reddening corrections narrow color--magnitude diagram (CMD) sequences in 369 of 435 clusters (85\%) with reliable CMD-width measurements, and cluster color excess maps reveal small-scale extinction structures. 
The median CER is compatible with the standard diffuse interstellar medium extinction curve, while the broad CER distribution and its large-scale variations across the Galactic disk likely reflect differences in the dominant dust environments sampled along different Galactic sight lines. 
\end{abstract}


\section{Introduction} 
Interstellar dust is a fundamental component of the Galactic interstellar medium and a major source of systematic uncertainty in astronomical observations. Dust extinction modifies observed magnitudes and colors over a wide wavelength range and directly affects open cluster (OC) studies \citep{2018SSRv..214...74M}. This is important for precision studies of OCs and the Galactic disk in the \textit{Gaia} era, because reddening and extinction uncertainties can introduce systematic biases rather than merely random errors. Differential reddening also broadens and distorts the color--magnitude diagram (CMD) sequence, weakens the constraining power of isochrone fitting, and increases the uncertainty of inferred cluster parameters \citep{2024A&A...686A..42H, 2024NewAR..9901696C, 2018A&A...618A..93C}. 

Previous studies have examined several aspects of extinction in OCs. Mean reddening and extinction have long been derived as basic parameters in cluster analyses \citep{2025AJ....170..288L, 2025A&A...703A.100N, 2024AJ....167...12C, 2019AJ....158..122K, 2020A&A...640A...1C}. Differential reddening has been recognized as an important source of CMD broadening and parameter uncertainty in individual clusters and in studies of limited cluster samples \citep{2024NewAR..9901696C, 2018A&A...618A..93C}. The wavelength dependence of extinction has also been investigated for OCs through color excess ratios (CERs) and related analyses of the extinction law \citep{2009ApJ...707..510Z, 2003A&A...397..191P, 2017PASA...34...68R}. However, these studies typically focus on one aspect of extinction or on relatively limited samples, and therefore do not yet provide a homogeneous, member-based framework in which mean reddening, differential reddening, and CERs are measured together for the same large OC sample. 

\textit{Gaia} astrometry and photometry now provide large, homogeneous samples of cluster members \citep{2024A&A...686A..42H, 2023A&A...673A.114H, 2021A&A...646A.104H, 2022A&A...661A.118C, 2020A&A...635A..45C, 2021MNRAS.504..356D, 2018A&A...618A..93C}. Combined with multi-band surveys such as \textit{2MASS} and large stellar parameter catalogs \citep{2023ApJS..267....8A, 2023MNRAS.524.1855Z, 2024MNRAS.52710937Y, 2024A&A...691A..98K}, these data make it possible to estimate intrinsic colors and reddening for individual members. 
Reddening estimates for member stars are especially useful in highly differentially reddened clusters, where a single mean reddening cannot recover the intrinsic CMD morphology. They also enable cluster-scale color excess maps, providing a finer view of spatial reddening variations within the member star fields than coarse three-dimensional dust maps in some cluster regions. 

In this work, we present homogeneous, member-based measurements of mean reddening, differential reddening, and CER for a large sample of Galactic OCs. 
Using \textit{Gaia} DR3 and \textit{2MASS} photometry together with stellar parameters from the SHBoost catalog \citep{2024A&A...691A..98K}, we derive color excesses for member stars and characterize extinction at the cluster scale. Based on these measurements, we investigate differential reddening and its impact on CMDs, construct cluster color excess maps, and examine how CER varies across the Galactic disk and what these variations imply for the Galactic extinction law. 
The paper is organized as follows. Section \ref{sec:data} describes the cluster sample, photometric data, and stellar parameters. Section \ref{sec:method} presents the derivation of color excesses for member stars and the reddening, differential reddening, and CER of each cluster. 
Section \ref{sec:res} presents the results of these measurements and discusses their implications.
Our conclusions are summarized in Section \ref{sec:summary}.

\section{Data and Sample} \label{sec:data}
To characterize the extinction properties of OCs, we require three basic ingredients: a homogeneous sample of cluster member stars, reliable multi-band photometry to measure observed colors, and stellar parameters to estimate intrinsic colors. 
This section describes the OC member sample, the optical and near-infrared photometry, and the adopted stellar parameter catalog.

\subsection{Open Cluster Sample} \label{sec:oc}
We adopted the membership catalog of 1481 OCs from \citet{2020AA...633A..99C}, which is built from \textit{Gaia} DR2 astrometry and photometry and provides, for each cluster, a list of candidate members with membership probabilities. Cluster memberships are determined with the UPMASK (Unsupervised Photometric Membership Assignment
in Stellar Clusters, \citet{2014A&A...561A..57K}) algorithm, which identifies overdensities in astrometric space (parallax and proper motions) and evaluates the spatial concentration on the sky through repeated resampling, yielding a membership probability for each star. Since \textit{Gaia}-based member determination may be incomplete for sparse, distant, or highly extincted clusters and can be more easily affected by field star contamination in crowded background regions, we applied an additional membership-probability threshold (80\%) to reduce possible systematic biases in the measured color residuals and CERs.

\subsection{Optical and Near-Infrared Photometric Data} \label{sec:photometric data}
The optical photometry, $G_{\rm BP}$ and $G_{\rm RP}$, was taken from \textit{Gaia} DR3 \citep{2023A&A...674A...1G}, while the near-infrared $K_{\rm S}$ photometry was adopted from the \textit{2MASS} Point Source Catalog \citep{2006AJ....131.1163S}. We cross-matched the \textit{Gaia} DR3 sources of OC member stars with \textit{2MASS} using a $1\arcsec$ search radius. To reduce photometric uncertainties while retaining enough stars for cluster CER fitting, we applied quality cuts to the \textit{Gaia} and \textit{2MASS} data. Specifically, we retained only stars with $K_{\rm S}$-band uncertainties smaller than 0.05 mag and $G_{\rm BP}$ and $G_{\rm RP}$ uncertainties smaller than 0.03 and 0.02 mag, respectively. We also required fractional parallax uncertainties smaller than $20\%$ and removed stars with $G>17$ mag or ${\rm RUWE}>1.4$.

We tested the sensitivity of the results to the adopted data quality selections by repeating the analysis after removing each criterion individually, including the photometric uncertainty cuts, the 80\% membership-probability threshold, the fractional parallax uncertainty cut, the $G<17$ magnitude limit, and the RUWE cut.
For each test, we compared three cluster-level quantities with the original measurements: the mean reddening, the differential reddening, and the CER.
The median differences are close to zero for all tests, indicating that the adopted selections do not introduce significant systematic biases. The largest effects arise from the $K_{\rm S}$-band uncertainties and membership-probability cuts, for which the median absolute changes reach $\sim0.05$ mag in reddening and differential reddening, and $\sim0.01$ in the CER. 
Lower-quality $K_{\rm S}$ photometry and lower-probability members mainly increase the dispersion in the color excesses of member stars. We therefore retain these cuts in the final analysis to suppress noisy measurements and improve the reddening and CER estimates.

\subsection{Stellar Parameters} \label{sec:stellar parameters}
\citet{2024A&A...691A..98K} derived stellar parameters from \textit{Gaia} DR3 XP spectra using a gradient-boosted random forest regressor (XGBoost). The model was trained on a dataset of approximately 7 million stars from high-quality parameters provided by the \texttt{StarHorse} code and major spectroscopic surveys. This approach enabled the estimation of reliable stellar parameters for about 217 million stars, including white dwarfs and hot stars that are often underrepresented in other surveys. The resulting SHBoost catalog provides estimates of effective temperature ($T_{\rm eff}$), surface gravity ($\log~g$), metallicity ([M/H]), and stellar mass, along with robust uncertainty estimates. 

We cross-matched the selected OC members with the SHBoost catalog using \textit{Gaia} \texttt{source\_id} to obtain stellar parameters ($T_{\rm eff}$, $\log~g$, and $[\rm M/H]$) for each star. To ensure parameter reliability, we applied quality cuts based on catalog-provided flags: uncertainties in $\log~g$ and [M/H] were required to be below 0.3, and the $\log T_{\rm eff}$ error below 0.1. We further limited the sample to stars with $T_{\rm eff}$ between 4000 and 8000 K and $\log~g$ between 1 and 5, corresponding to the well calibrated regime of the SHBoost model \citep[see][]{2024A&A...691A..98K}. Outside this range, especially at $T_{\rm eff} \lesssim 4000$ K and $\gtrsim 8000$ K, the model has reduced generalization and larger uncertainties.

\section{Method} \label{sec:method}
\subsection{Color Excess Measurements for Member Stars}\label{sec:color_excess}

To derive the mean reddening, differential reddening, and CER of each OC, we first measured $E(G_{\rm BP}-G_{\rm RP})$ and $E(G_{\rm BP}-K_{\rm S})$ for its member stars. For each star, the color excess is the observed color minus the intrinsic color. The observed colors, $(G_{\rm BP}-G_{\rm RP})$ and $(G_{\rm BP}-K_{\rm S})$, come directly from \textit{Gaia} and \textit{2MASS} photometry, so we required estimates of the intrinsic colors.

We adopted the ``blue-edge'' method, a widely used approach to estimate stellar intrinsic color indices \citep{2014ApJ...788L..12W, 2019ApJ...877..116W, 2023ApJ...956...26L, 2023ApJ...946...43W, 2025ApJ...982...77D}. This method establishes empirical relations between $T_{\rm eff}$ (and, in some implementations, [M/H]) and intrinsic colors for a given luminosity class. To improve the generalization of this approach, \citet{2024ApJ...974..138Z} trained an XGBoost model on over one million low reddening stars, using \textit{Gaia} DR3 GSP-Phot parameters ($T_{\rm eff}$, $\log~g$, [M/H]) as input. The model predicts intrinsic color indices in both \textit{Gaia} and \textit{2MASS} bands, including $(G_{\rm BP}-G_{\rm RP})_0$, $(G_{\rm BP}-K_{\rm S})_0$, and $(J-K_{\rm S})_0$.
In this work, we applied this model to SHBoost stellar parameters \citep{2024A&A...691A..98K} to estimate $(G_{\rm BP}-G_{\rm RP})_0$ and $(G_{\rm BP}-K_{\rm S})_0$ for OC member stars. 
The XGBoost intrinsic color model was trained using GSP-Phot parameters, whereas we used SHBoost parameters. Nevertheless, both parameter sets are derived from \textit{Gaia} DR3 XP spectra.
\citet{2024ApJ...974..138Z} showed that the predicted intrinsic colors are only weakly affected by moderate systematic offsets in the input stellar parameters. They also obtained consistent reddening estimates when applying the model to the independent stellar parameter catalog of \citet{2023ApJS..267....8A}. 
We therefore treated the adopted stellar parameter scale as a separate uncertainty source and evaluated its impact on the cluster measurements at the end of this section.

For each OC member star, we calculated the color excesses as 
$E(G_{\rm BP}-G_{\rm RP}) = (G_{\rm BP} - G_{\rm RP}) - (G_{\rm BP}-G_{\rm RP})_0$
and
$E(G_{\rm BP}-K_{\rm S}) = (G_{\rm BP} - K_{\rm S}) - (G_{\rm BP}-K_{\rm S})_0$. 
Because \textit{Gaia} XP spectra have relatively low spectral resolution, the stellar parameters inferred from them, and hence the predicted intrinsic colors, are model dependent. Their uncertainties therefore need to be propagated into the color excesses. To estimate the statistical uncertainty in the color excess of each star, we propagated the SHBoost uncertainties in $T_{\rm eff}$, $\log~g$, and metallicity through the XGBoost intrinsic color model. This gives an uncertainty term, $\sigma_{{\rm par},i}$, that reflects the effect of uncertainties in the stellar parameters on the predicted intrinsic color. 
We also included the representative uncertainties in the predicted intrinsic colors reported by \citet{2024ApJ...974..138Z}: $\sigma_{\rm IC}=0.014$~mag for $(G_{\rm BP}-G_{\rm RP})_0$ and $\sigma_{\rm IC}=0.050$~mag for $(G_{\rm BP}-K_{\rm S})_0$. These values represent the residual scatter between the predicted and reference intrinsic colors in an independent low reddening test sample. For each color index, we combined these terms in quadrature with the corresponding photometric uncertainty, $\sigma_{{\rm phot},i}$, to obtain the color excess uncertainty for each star: 
$\sigma_{E_i}^2=\sigma_{{\rm phot},i}^2+\sigma_{{\rm par},i}^2+\sigma_{\rm IC}^2.$ 
We then propagated these uncertainties into the statistical uncertainties of the cluster measurements. The resulting color excesses are used to measure the reddening, differential reddening, and CER of each cluster. 
Using this procedure, we identified 1225 OCs whose member stars have measurable reddening.

\subsection{Mean Reddening and Differential Reddening of Open Clusters}\label{sec:mean_dr}

For a given OC, the member stars occupy a relatively small area on the sky and lie at approximately the same distance. Their line of sight extinction is therefore expected to be broadly similar. We used the mean color excess of the member stars to represent the cluster reddening ($E$), that is, the reddening along the line of sight to the cluster at its distance.
For each cluster, we estimated the mean color excess using a maximum-likelihood model. The model accounts for the color excess uncertainty of each star, $\sigma_{E_i}$, and an additional dispersion term. This term represents residual scatter among member stars that cannot be explained by their measurement uncertainties. We included it when estimating the statistical uncertainty of the mean reddening, $\sigma_E^{\rm stat}$, to avoid underestimating this uncertainty when differential reddening or other residual scatter is significant. We did not use this dispersion term as a measure of differential reddening. Instead, we quantified differential reddening separately using the MAD defined below. 

We used the spread in the color excesses of member stars as an observational measure of differential reddening within each cluster. It mainly reflects inhomogeneous extinction across the cluster field, but may also include contributions from photometric uncertainties, intrinsic color errors, and residual field star contamination. The high membership-probability threshold and the quality cuts on photometry and stellar parameters reduce these non-physical contributions.
We quantified differential reddening using the median absolute deviation (MAD). For each cluster, we defined
\begin{equation}\label{eq:mad}
{\rm MAD}(E) = {\rm median}\left( \left| E_i - {\rm median}(E_i) \right| \right),
\end{equation}
where $E_i$ is the color excess of the $i$th member star. 
We estimated the statistical uncertainty of the MAD, $\sigma_{\rm MAD}^{\rm stat}$, with bootstrap resampling. In each realization, member stars were drawn with replacement, the color excess of each selected star was perturbed according to its $\sigma_{E_i}$ value, and the MAD was recalculated. We then adopted half of the 16th--84th percentile interval of the bootstrap distribution as $\sigma_{\rm MAD}^{\rm stat}$.
We applied this definition to $E(G_{\rm BP}-G_{\rm RP})$ and $E(G_{\rm BP}-K_{\rm S})$, hereafter denoted as ${\rm MAD}(E_{\rm BP-RP})$ and ${\rm MAD}(E_{\rm BP-K_{\rm S}})$.

\subsection{Color Excess Ratios of Open Clusters}

The CER is widely used to characterize the extinction law and can be related to quantities such as $R_{\rm V}$ and the dust grain size distribution \citep{2009ApJ...707..510Z, 2016ApJ...821...78S, 2023ApJ...956...26L}. Because the CER is the ratio of two color excesses, it largely removes the dependence on the absolute dust column and instead characterizes the relative wavelength dependence of extinction. For each OC, we determined the CER as the slope of the color excess relation. Specifically, we performed a weighted linear fit of $E(G_{\rm BP}-K_{\rm S})$ against $E(G_{\rm BP}-G_{\rm RP})$, constrained to pass through the origin and accounting for the uncertainties in both color excesses. The fitted slope $k_{\rm oc}$ was adopted as the cluster CER. The zero-intercept constraint is physically motivated by the fact that both color excesses vanish in the absence of reddening. 
We also tested a CER fit with a free intercept. Across the sample, the median fitted intercept is $b=0.026$ mag, smaller than the typical uncertainty $\sigma_b=0.068$ mag, with a median significance of $b/\sigma_b=0.406$. In addition, 78\% of the clusters have intercepts consistent with zero within $2\sigma$. This indicates no significant systematic zero-point offset and supports the adopted zero-intercept model.

We estimated the statistical uncertainty of the CER, $\sigma_{k}^{\rm stat}$, with the same bootstrap approach. In each realization, member stars were drawn with replacement, and the two color excesses of each selected star were perturbed according to their respective uncertainties. The weighted linear fit constrained to pass through the origin was then repeated, and $\sigma_k^{\rm stat}$ was taken as half of the 16th--84th percentile interval of the resulting $k_{\rm oc}$ distribution. We defined a signal-to-noise proxy as ${\rm SNR}=k_{\rm oc}/\sigma_k^{\rm stat}$ and computed the coefficient of determination $R^2$ to characterize the fit. The SNR quantifies the constraint on the derived CER, while $R^2$ evaluates the tightness of the linear relation between the two color excesses. Higher SNR and $R^2$ values indicate more reliable CER measurements.

The derived color excesses may be affected by binary systems, since neither the SHBoost stellar parameters nor the intrinsic color model explicitly accounts for them. To evaluate their impact, we cross-matched our member star sample with the \textit{Gaia} DR3 non-single star (NSS) catalog and repeated the full analysis after excluding all \textit{Gaia}-identified NSS sources. In total, 169 stars (0.36\%) were removed from the original sample. 
For each cluster, the resulting changes in reddening, differential reddening, and CER are negligible: the median signed differences are consistent with zero. For more than 99\% of clusters, these changes are smaller than the corresponding statistical uncertainties. These results indicate that \textit{Gaia}-identified NSS sources are not a significant source of systematic uncertainty in our measurements.

We then evaluated the systematic uncertainties associated with the adopted stellar parameter scale. Because the reported quantities are cluster measurements derived from member stars, we assessed the effect of this scale by repeating the full analysis with alternative input parameters. 
We first recalculated the intrinsic colors using the SHBoost and GSP-Phot parameter sets separately, and then rederived the mean reddening, differential reddening, and CER of each cluster. The absolute difference between the results obtained with the two parameter sets was adopted as the systematic uncertainty associated with the stellar parameter scale, $\sigma_{\rm sys,par}$. We also tested a possible zero-point offset in the SHBoost $T_{\rm eff}$ scale by shifting the temperatures by $\pm100$ K and repeating the analysis. 
For each cluster, we used the root-mean-square change induced by the $+100$ and $-100$ K shifts relative to the result obtained with the unshifted SHBoost temperatures, $\sigma_{{\rm sys},T_{\rm eff}}=[(\Delta_{+100}^{2}+\Delta_{-100}^{2})/2]^{1/2}$, as the temperature scale contribution.
The total systematic uncertainty was obtained by combining these two contributions in quadrature, i.e., $\sigma_{\rm sys}^{2}=\sigma_{{\rm sys},par}^{2}+\sigma_{{\rm sys},T_{\rm eff}}^{2}$. The systematic uncertainties ($\sigma_{\rm sys}$) are reported separately from the statistical uncertainties in Table~\ref{tab:table1}.

\begin{deluxetable*}{lcccccccccc}
\tabletypesize{\scriptsize}
\setlength{\tabcolsep}{3pt}
\tablewidth{0pt}
\tablecaption{Reddening, Differential reddening, and CERs of OCs\label{tab:table1}}
\tablehead{
\colhead{Cluster} & \colhead{RA}    & \colhead{DEC}   & \colhead{distance}     & \colhead{$E(G_{\rm BP}-G_{\rm RP})$}   &  \colhead{$E(G_{\rm BP}-K_{\rm S})$} & \colhead{$A_{\rm V}^{a}$} & \colhead{${\rm MAD}(E_{\rm BP-RP})$} & \colhead{${\rm MAD}(E_{\rm BP-K_{\rm S}})$} & \colhead{$\frac{E(G_{\rm BP}-K_{\rm S})}{E(G_{\rm BP}-G_{\rm RP})}$} & \colhead{$N_{\rm used}^{b}$}\\
\colhead{}         & \colhead{(deg)} & \colhead{(deg)} & \colhead{(pc)} & \colhead{(mag)}  & \colhead{(mag)}               & \colhead{(mag)}      & \colhead{(mag)}                & \colhead{(mag)}       & \colhead{}    & \colhead{}
}
\startdata
ASCC~10  & 51.870 & 34.981 & 672.0  &
$0.303\pm0.008\pm0.036$ & $0.657\pm0.020\pm0.080$ &
$0.726\pm0.019\pm0.086$ & $0.048\pm0.013\pm0.006$ & $0.149\pm0.036\pm0.016$ & $2.178\pm0.024\pm0.022$ & 33 \\
ASCC~101 & 288.399 & 36.369 & 397.3  &
$0.053\pm0.006\pm0.040$ & $0.123\pm0.016\pm0.083$ &
$0.128\pm0.013\pm0.097$ & $0.042\pm0.012\pm0.015$ &
$0.111\pm0.033\pm0.032$ & $2.281\pm0.153\pm0.155$ & 36 \\
ASCC~105 & 295.548 & 27.366 & 551.8  &
$0.174\pm0.008\pm0.037$ & $0.404\pm0.019\pm0.085$ &
$0.416\pm0.019\pm0.090$ & $0.059\pm0.011\pm0.004$ &
$0.133\pm0.031\pm0.039$ & $2.304\pm0.036\pm0.046$ & 64 \\
ASCC~107 & 297.164 & 21.987 & 878.5  & $0.700\pm0.035\pm0.031$ & $1.442\pm0.071\pm0.070$ & $1.675\pm0.084\pm0.075$ & $0.158\pm0.044\pm0.016$ & $0.377\pm0.091\pm0.036$ & $2.043\pm0.018\pm0.016$ & 35 \\
ASCC~108 & 298.306 & 39.349 & 1154.0 & $0.211\pm0.007\pm0.028$ & $0.441\pm0.015\pm0.064$ & $0.505\pm0.016\pm0.066$ & $0.051\pm0.009\pm0.007$ & $0.102\pm0.020\pm0.020$ & $2.075\pm0.022\pm0.042$ & 93 \\
ASCC~11  & 53.056 & 44.856 & 854.5  & $0.368\pm0.007\pm0.030$ & $0.772\pm0.014\pm0.069$ & $0.880\pm0.016\pm0.072$ & $0.067\pm0.009\pm0.005$ & $0.167\pm0.022\pm0.005$ & $2.081\pm0.009\pm0.028$ & 145 \\
ASCC~110 & 300.742 & 33.528 & 1902.2 & $0.503\pm0.026\pm0.030$ & $1.090\pm0.061\pm0.062$ &
$1.203\pm0.063\pm0.073$ & $0.087\pm0.028\pm0.007$ &
$0.245\pm0.076\pm0.010$ & $2.151\pm0.038\pm0.013$ & 32 \\
ASCC~111 & 302.891 & 37.515 & 836.9  & $0.297\pm0.010\pm0.030$ & $0.652\pm0.023\pm0.068$ &
$0.711\pm0.024\pm0.073$ & $0.065\pm0.014\pm0.006$ &
$0.134\pm0.035\pm0.014$ & $2.194\pm0.029\pm0.025$ & 54 \\
ASCC~114 & 324.990 & 53.997 & 913.2  &
$0.451\pm0.009\pm0.030$ & $0.998\pm0.021\pm0.069$ &
$1.080\pm0.021\pm0.072$ & $0.066\pm0.013\pm0.007$ &
$0.160\pm0.031\pm0.017$ & $2.208\pm0.017\pm0.020$ & 66 \\
ASCC~115 & 329.280 & 51.558 & 746.0  &
$0.392\pm0.019\pm0.027$ & $0.849\pm0.043\pm0.061$ &
$0.938\pm0.045\pm0.064$ & $0.102\pm0.025\pm0.005$ &
$0.190\pm0.060\pm0.024$ & $2.161\pm0.021\pm0.022$ & 16 \\
... & ... & ... & ... & ... & ... & ... & ... & ... & ... & ... \\
\enddata
Notes: \\
For the reddening-related quantities, the first and second quoted uncertainties represent the statistical ($\sigma_{\rm stat}$) and systematic ($\sigma_{\rm sys}$) uncertainties, respectively. Their estimation is described in \ref{sec:method}. The additional propagation for $A_{\rm V}$ is given in note a.

$\rm ^a$ $A_{\rm V}$ is computed from the mean cluster reddening using
$A_{\rm V} = (2.394 \pm 0.018)\,E(G_{\rm BP}-G_{\rm RP})$
following \citet{2019ApJ...877..116W}. 
The statistical and systematic uncertainties in $A_{\rm V}$ are propagated separately from the corresponding uncertainties in $E(G_{\rm BP}-G_{\rm RP})$, while the uncertainty in the conversion coefficient is included in the systematic term. 
We note that this fixed conversion factor represents an average Galactic extinction relation and is adopted independently of the fitted relations derived in Figures~\ref{fig:2}(b) and \ref{fig:3}.

$\rm ^b$ $N_{\rm used}$: Number of member stars used for fitting after quality cuts.

Cluster parameters are adopted from \citet{2020AA...633A..99C}, reddening related quantities are derived in this work.

(This table is available in its entirety in machine readable form.) \\
\end{deluxetable*}

\begin{figure}[!htbp]
\centering
\includegraphics[scale=0.7]{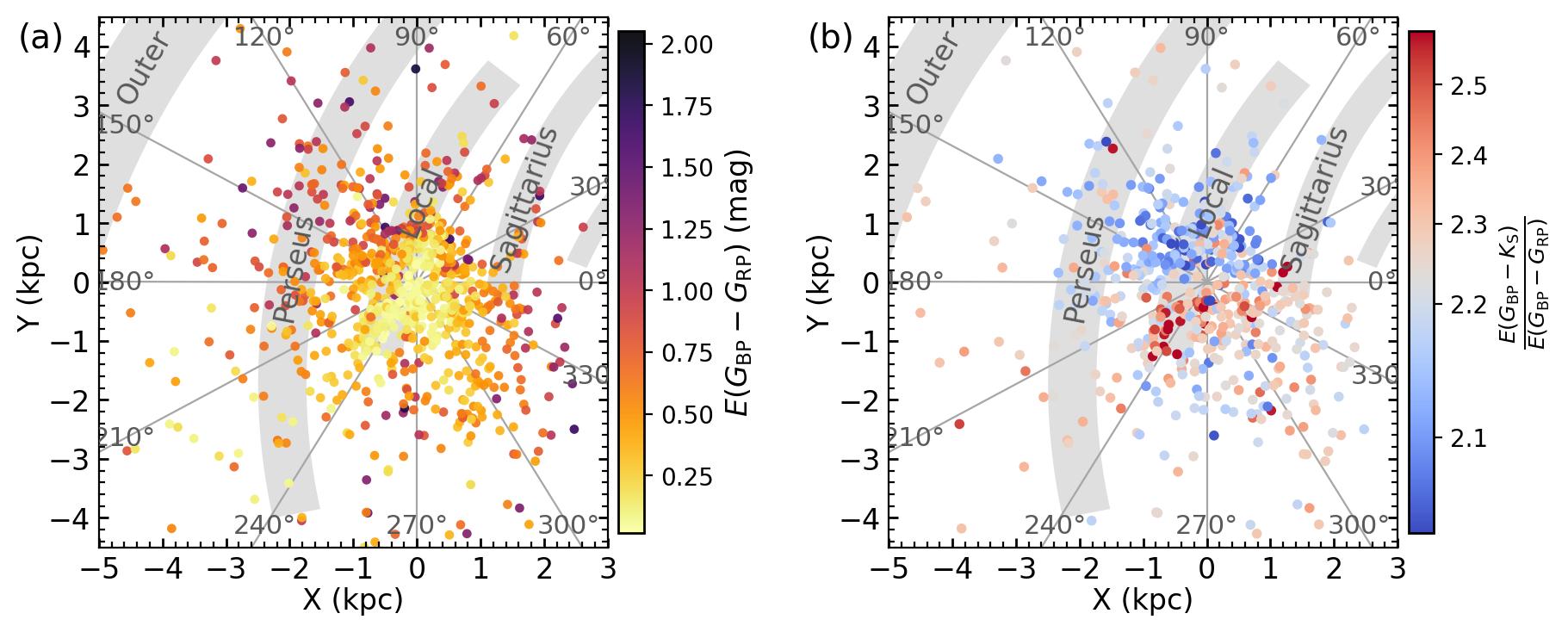}
\figurenum{1}
\caption{Spatial distribution of the OCs in the Galactic $X$--$Y$ plane. Panel (a) is color coded by the mean reddening $E(G_{\rm BP}-G_{\rm RP})$, while panel (b) is color coded by the CER. The top-down projection of the Galactic disk is overlaid with lines of constant Galactic longitude to guide the eye in azimuth. The gray shaded regions indicate the approximate locations of the Galactic spiral arms adopted from the model of \citet{2014ApJ...783..130R}. In panel (b), the color scale is centered on the fiducial CER value $k_{3.1}=2.22$, corresponding to the standard extinction law with $R_{\rm V}=3.1$.
}
\label{fig:1}
\end{figure}

\section{Results and Discussion} \label{sec:res}
\subsection{Distribution of Cluster Reddening} \label{sec:4.1}

Following the procedures described in Section \ref{sec:method}, we derive the mean reddening, differential reddening, and CER for the OCs. Table~\ref{tab:table1} illustrates the catalog format with 10 representative entries. The complete table is available in machine readable form. 
For reliable cluster statistics, we retain only clusters with color excess measurements for more than ten member stars ($N_{\rm used}\geq10$), leaving 729 OCs for the color excess analysis. Their mean $E(G_{\rm BP}-G_{\rm RP})$ and $E(G_{\rm BP}-K_{\rm S})$ values are 0.471 and 1.049 mag, respectively. Figure \ref{fig:1}(a) shows the distribution of $E(G_{\rm BP}-G_{\rm RP})$ in the Galactic $X$--$Y$ plane. Reddening ranges from $\lesssim 0.2$ mag to $\gtrsim 1.5$--2 mag and is spatially non-uniform across the Galactic plane. The spiral arm loci in Figure \ref{fig:1} are used only as a large-scale reference. The sample spans a broad age range and is not expected to trace the spiral pattern one-to-one: young OCs are more concentrated along nearby spiral arm features, whereas older OCs are more dispersed throughout the Galactic disk \citep{2020A&A...640A...1C}. Moreover, the large-scale spiral structure of the Milky Way remains uncertain because different tracers and methods yield different morphologies \citep{2018RAA....18..146X, 2021A&A...652A.102H}.

\subsection{Comparison with Independent Reddening or Extinction Estimates}\label{sec:4.2}

\begin{figure}[!htbp]
\centering
\includegraphics[width=\textwidth]{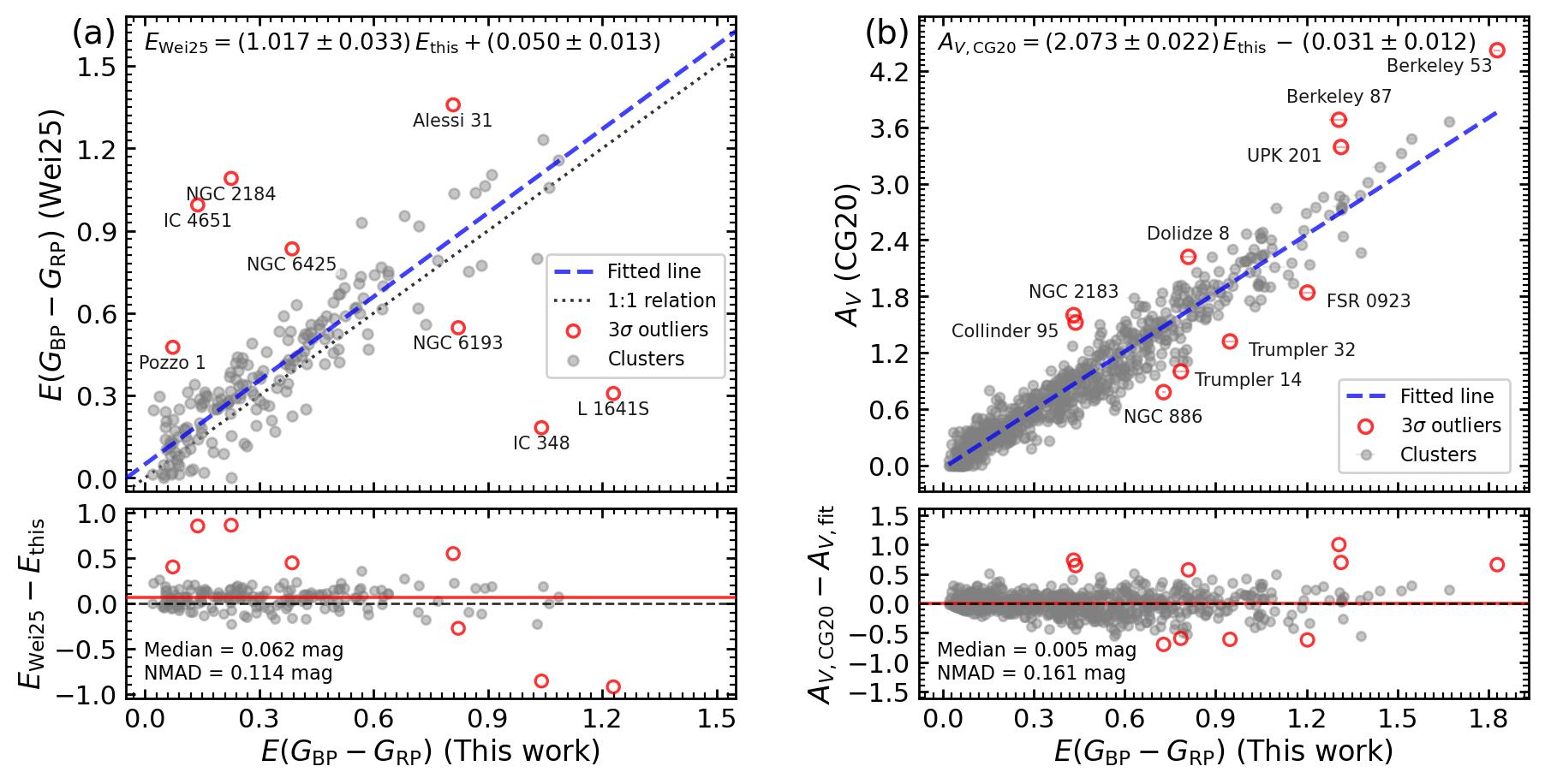}
\figurenum{2}
\caption{External comparisons of the OC reddening and extinction measurements. Panel~(a) compares the $E(G_{\rm BP}-G_{\rm RP})$ values derived in this work with those of \citet{2025AJ....169..115W} for the 186 clusters in common. The blue dashed line shows the best fitting relation after iterative $3\sigma$ clipping, the black dotted line marks the one-to-one relation, and red open circles indicate the identified $3\sigma$ outliers. Panel~(b) compares our $E(G_{\rm BP}-G_{\rm RP})$ measurements with the $A_{\rm V}$ values of \citet{2020A&A...640A...1C}. The blue dashed line shows the best fitting relation, red open circles mark clusters deviating by more than $3\sigma$. In both lower subpanels, the red solid line indicates the median residual and the black dashed line marks zero.}
\label{fig:2}
\end{figure}

We first compare our cluster mean reddening with two independent estimates of OC reddening or extinction. Figure~\ref{fig:2}(a) compares our results in the same \textit{Gaia} color index with \citet{2025AJ....169..115W}, who derived cluster reddening from \textit{Gaia} CMD isochrone fitting. Among the 280 clusters analyzed by \citet{2025AJ....169..115W}, 186 have reliable $E(G_{\rm BP}-G_{\rm RP})$ measurements in our sample ($N_{\rm used}\geq10$) and are used for the comparison. The two datasets are broadly consistent: the median difference is $\sim$0.062 mag, with a normalized median absolute deviation (NMAD) scatter of $\sim$0.114 mag. A linear fit gives a slope consistent with unity, indicating no significant difference in the reddening scale between the two studies, although a small positive zeropoint offset of about 0.05 mag is present.
For several outliers, our reddening estimates are closer to previous literature values than those of \citet{2025AJ....169..115W}. These differences may arise from membership selection and CMD morphology, which they also identified as sources of discrepancies in their comparisons with earlier studies \citep{2020A&A...640A...1C, 2021MNRAS.504..356D, 2023A&A...673A.114H, 2024AJ....167...12C}.

Figure~\ref{fig:2}(b) compares our mean reddening $E(G_{\rm BP}-G_{\rm RP})$ with the $A_{\rm V}$ values of \citet{2020A&A...640A...1C} for all 729 clusters with valid reddening measurements. They used an artificial neural network applied to \textit{Gaia} CMDs and median parallaxes to estimate cluster $A_{\rm V}$. The two quantities are positively correlated, with a fitted slope of $2.073\pm0.022$. This is smaller than the value expected for a standard $R_{\rm V}=3.1$ extinction curve of $2.394\pm0.018$ \citep{2019ApJ...877..116W}. As discussed in Section~\ref{sec:4.4}, the observed median CER is compatible with, but does not uniquely determine, the standard extinction curve. We therefore interpret the slope difference mainly as a systematic offset between different extinction estimation methods, rather than as evidence for a strongly non-standard mean extinction law. For the subsequent analysis, we adopt the literature conversion factor of $2.394\pm0.018$ to convert $E(G_{\rm BP}-G_{\rm RP})$ to $A_{\rm V}$, and list the resulting $A_{\rm V}$ values in Table~\ref{tab:table1}.

These comparisons indicate a broadly consistent reddening scale with method-dependent scatter. Our method derives star-by-star color excesses from observed colors and intrinsic colors predicted from stellar parameters using a blue-edge-calibrated XGBoost model. CMD- or isochrone-based estimates instead depend on the adopted cluster sequence or locus and are sensitive to CMD morphology. Differential reddening can broaden the CMD sequence and increase the scatter in CMD-based extinction estimates. As illustrated in Figure~\ref{fig:2}(b), the outlying clusters generally show strong differential reddening (mean ${\rm MAD}(E_{\rm BP-RP})=0.19$~mag) and broader CMD sequences. In isochrone-based analyses, fitting the blue edge rather than the mean locus can reduce the influence of unresolved binaries because binaries tend to lie redward and/or brighter than the single star sequence. Our \textit{Gaia} NSS-removal test indicates that identified non-single stars do not significantly affect the reddening or differential reddening measurements for each cluster. In Section~\ref{sec:4.3}, we further assess the effect of differential reddening correction on the cluster CMDs.

\begin{figure}[!htbp]
\centering
\includegraphics[scale=0.9]{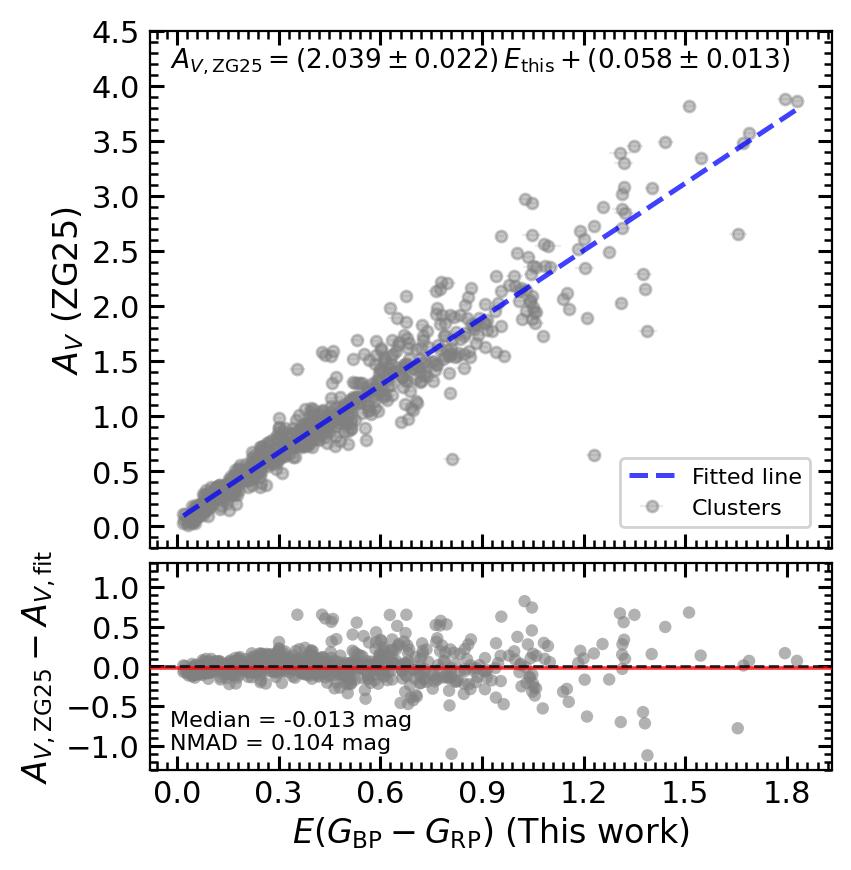}
\figurenum{3}
\caption{Comparison between the $A_{\rm V}$ values from the three-dimensional dust map of \citet{2025Sci...387.1209Z} at the cluster positions and distances and the cluster mean reddening $E(G_{\rm BP}-G_{\rm RP})$ derived in this work. Gray points denote individual OCs within 5 kpc. 
The blue dashed line shows the best fitting relation. In the lower panel, the red solid line indicates the median residual and the black dashed line marks zero.}
\label{fig:3}
\end{figure}

\begin{figure}[!htbp]
\centering
\includegraphics[width=\textwidth]{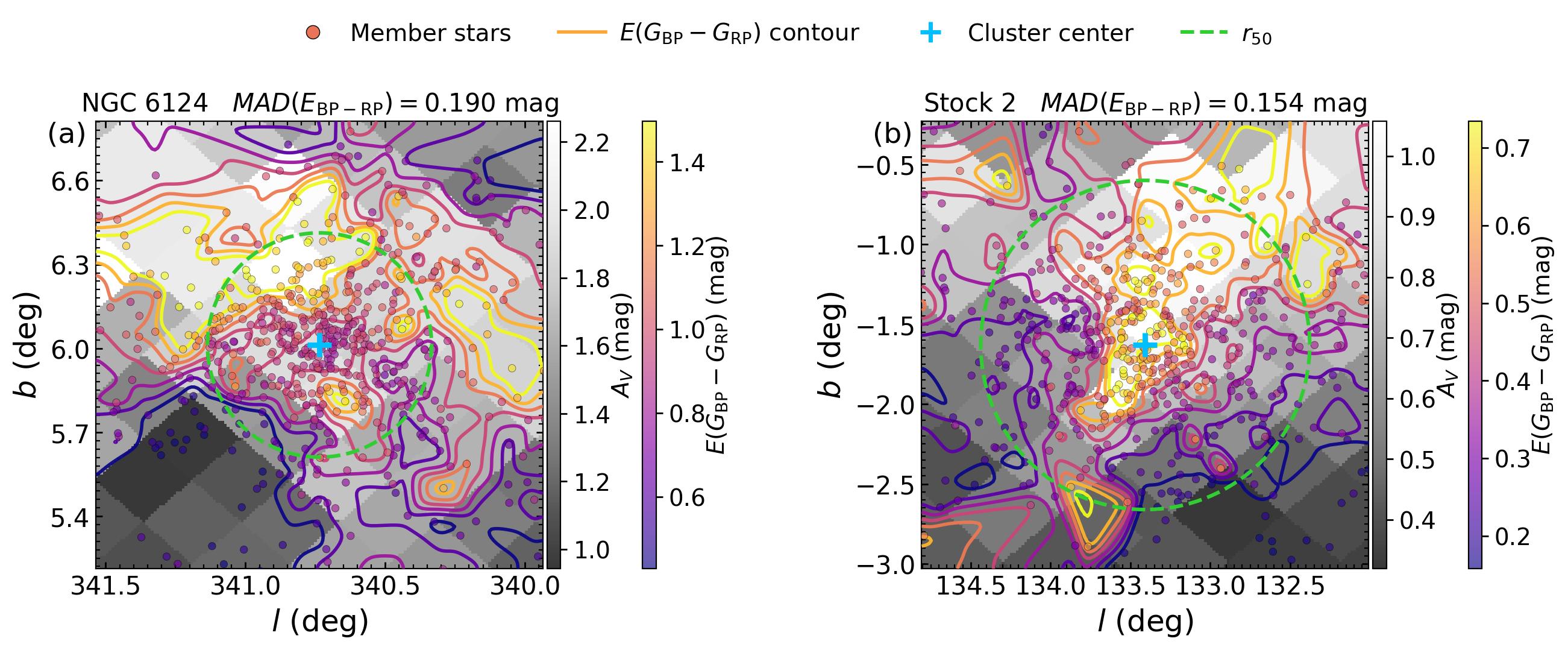}
\figurenum{4}
\caption{Comparison between the large-scale extinction structure traced by the dust map of \citet{2025Sci...387.1209Z} and the small-scale reddening variations revealed by individual cluster members. The gray background shows the cumulative $A_{\rm V}$ distribution from the dust map at the cluster distance, while the colored points indicate the measured $E(G_{\rm BP}-G_{\rm RP})$ values of member stars. The colored contours trace the smoothed spatial distribution of reddening among the member stars. The blue cross marks the cluster center and the dashed circle denotes $r_{50}$. 
Both clusters exhibit relatively large differential reddening, and the reddening distribution of the member stars reveals localized extinction structures that are only weakly reflected in the lower-resolution dust map.}
\label{fig:4}
\end{figure}

We also compare our cluster color excesses with the $A_{\rm V}$ values inferred from the 3D dust map of \citet{2025Sci...387.1209Z}. The map reconstructs the cumulative extinction as a function of distance from extinction increments between adjacent distance bins, which are converted to $A_{\rm V}$ and integrated along the line of sight. 
For each cluster, we extract this cumulative $A_{\rm V}$ at the corresponding sky position and cluster distance, and compare it with our $E(G_{\rm BP}-G_{\rm RP})$. Because the map is primarily constrained within the nearby few-kpc volume, we restrict the comparison to OCs within $5\,\mathrm{kpc}$.

As shown in Figure~\ref{fig:3}, the map-based $A_{\rm V}$ values are positively correlated with our reddening measurements, indicating that the two approaches recover the same broad variation in extinction. The dispersion of the residuals increases toward larger $E(G_{\rm BP}-G_{\rm RP})$. Beyond this reddening dependence, differences between the angular extent and spatial sampling of the cluster and the angular resolution and effective sampling scale of the dust map may further increase the residual dispersion. 
Figure~\ref{fig:4} illustrates this difference. The dust map reproduces the broad extinction pattern across the cluster fields, whereas the reddening measurements of individual member stars reveal localized variations that are smoothed in the map. Differential reddening maps for individual clusters are therefore better suited to characterizing the small-scale extinction structure relevant to individual cluster analyses.

The comparison with the three-dimensional dust map highlights the advantage of the member-based approach: it is tied to compact stellar systems at well defined distances and retains cluster-scale reddening structure that is smoothed out in coarser maps. This provides the basis for the differential reddening analysis presented below.

\subsection{Differential Reddening}\label{sec:4.3}

\begin{figure}[!htbp]
\centering
\includegraphics[scale=0.9]{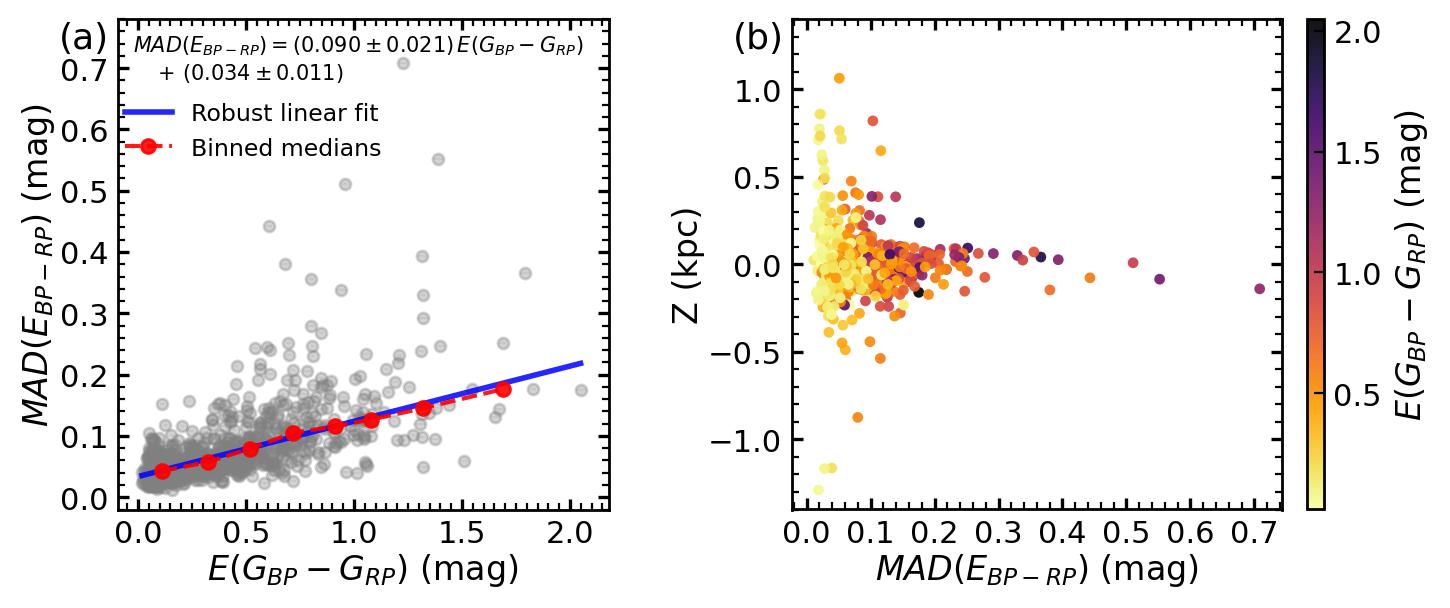}
\figurenum{5}
\caption{Differential reddening of the OC sample. Panel~(a) shows the cluster differential reddening, quantified by ${\rm MAD}(E_{\rm BP-RP})$, as a function of the mean reddening $E(G_{\rm BP}-G_{\rm RP})$. The blue solid line marks the Huber robust linear fit, while the red dots represent the median values in reddening bins. Panel~(b) shows ${\rm MAD}(E_{\rm BP-RP})$ as a function of Galactic height $Z$, with the points color coded by the mean reddening.}
\label{fig:5}
\end{figure}

We first examine the amplitude and spatial dependence of differential reddening across the OC sample. We then use the change in CMD-width after star-by-star dereddening as an empirical diagnostic of its contribution to the observed CMD broadening.
Figure~\ref{fig:5}(a) shows that differential reddening increases with mean reddening.
To reduce the influence of clusters with large differential reddening, we characterize the trend using a Huber robust linear fit, obtaining $\mathrm{MAD}(E_{\rm BP-RP}) = (0.090 \pm 0.021)\times E(G_{\rm BP}-G_{\rm RP}) + (0.034 \pm 0.011)$. The coefficient uncertainties were estimated from bootstrap resampling in reddening bins, accounting for the highly uneven distribution of clusters in mean reddening. The median values in reddening bins closely follow this relation, indicating that the correlation is not driven by a small number of extreme outliers. More highly reddened clusters therefore tend to exhibit a larger star-to-star spread in color excess. We use this relation only as a statistical description of the sample trend, not as a precise physical law for individual clusters.
Figure~\ref{fig:5}(b) shows differential reddening versus Galactic height $Z$, color coded by mean reddening. Low reddening clusters have weak differential reddening across a wide range of $|Z|$, whereas more highly reddened clusters are concentrated at low $|Z|$ and span a broader range of differential reddening. The median MAD decreases with distance from the Galactic plane, approaching $\sim0.047$ mag at $|Z| \gtrsim 200$ pc. These patterns suggest that clusters near the Galactic plane more often lie along sight lines with both larger dust columns and stronger small-scale dust inhomogeneity.

\begin{figure}[!htbp]
\centering
\includegraphics[width=\textwidth]{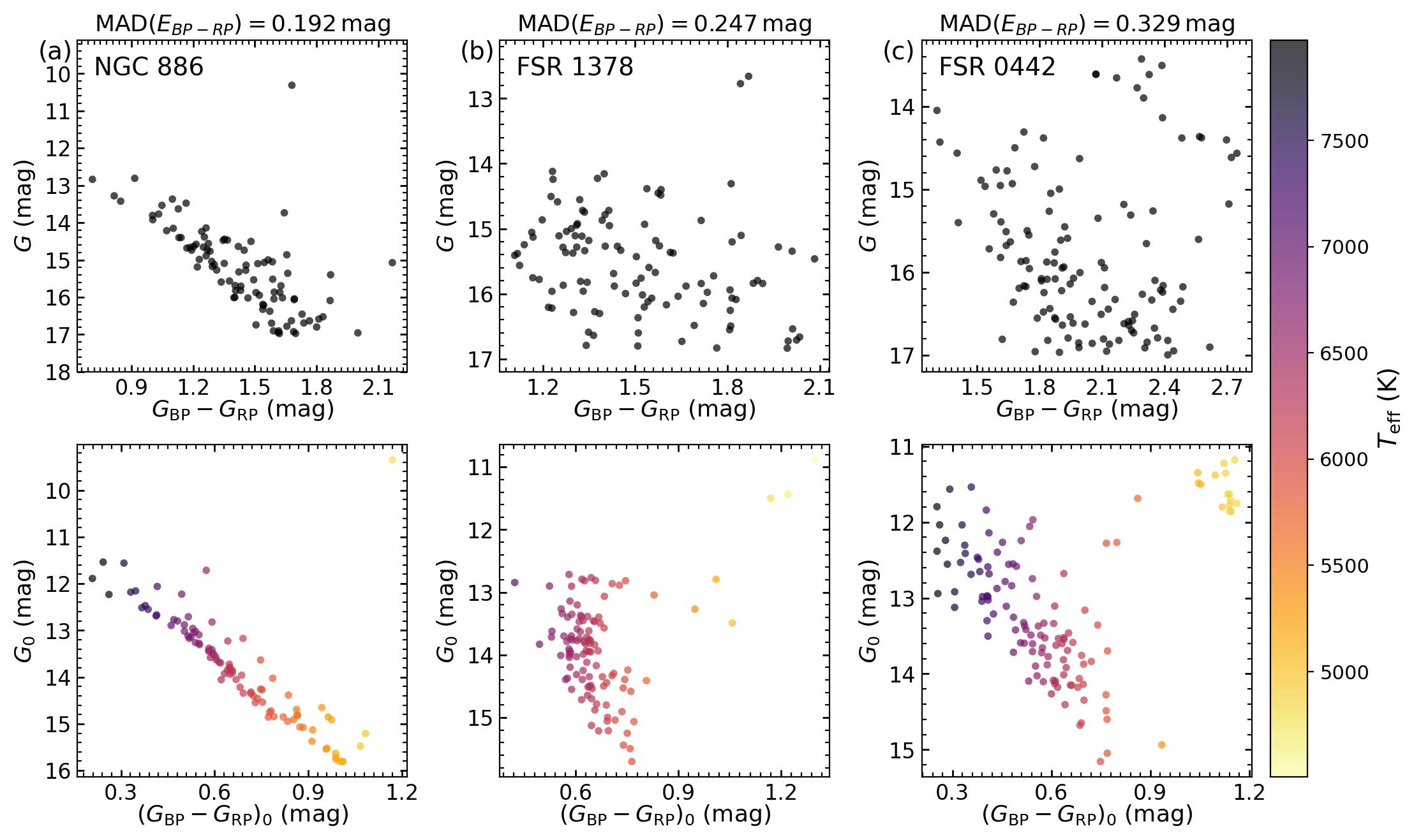}
\figurenum{6}
\caption{Examples of CMDs before and after differential reddening correction for three OCs. The upper panels show the observed CMDs ($G$ versus $G_{\rm BP}-G_{\rm RP}$), while the lower panels show the CMDs after correcting individual member stars for reddening and extinction using the color excesses derived in this work. For each cluster, the upper and lower panels use the same member star sample. In the lower panels, points are color coded by $T_{\rm eff}$. Comparing the upper and lower panels illustrates the impact of differential reddening on the observed CMD morphology.}
\label{fig:6}
\end{figure}

\begin{figure}[!htbp]
\centering
\includegraphics[scale=0.9]{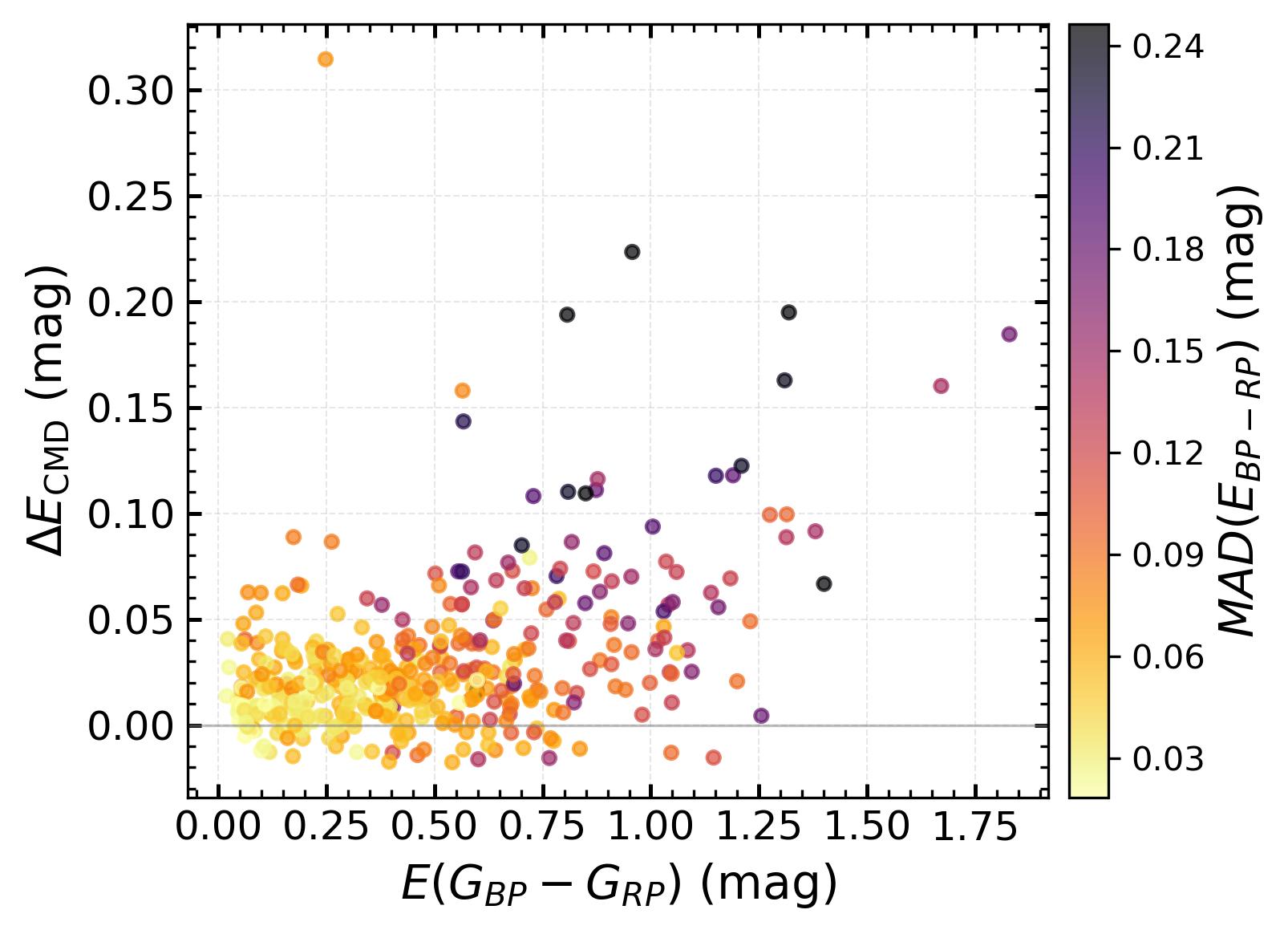}
\figurenum{7}
\caption{Change in CMD-width after differential reddening correction as a function of the mean color excess $E(G_{\rm BP}-G_{\rm RP})$.  Each point represents one open cluster, and the color scale indicates the differential reddening measured by $\mathrm{MAD}(E_{\rm BP-RP})$. Positive values of $\Delta E_{\rm CMD}$ correspond to a reduction in the CMD-width after dereddening, while the horizontal line marks $\Delta E_{\rm CMD}=0$.}
\label{fig:7}
\end{figure}

Differential reddening can broaden the cluster sequence and affect CMD-based analyses. We therefore use the change in CMD-width after star-by-star correction as an empirical check of how the measured reddening variations affect the observed CMD broadening.
Figure~\ref{fig:6} presents three representative examples. All three clusters have been included and characterized in multiple Gaia-based studies \citep{2024A&A...686A..42H, 2024AJ....167...12C, 2021MNRAS.504..356D}. 
The Unified Cluster Catalog \citep[UCC;][]{2023MNRAS.526.4107P} also classifies all three as rich, high-quality clusters. 
 
For each example, the observed and dereddened CMDs are constructed from the same stars with available color excess estimates for individual stars.
After dereddening, the stellar sequences generally become narrower and better defined, indicating that the measured reddening variations account for part of the observed CMD broadening.
To quantify this effect, we binned each CMD in $G$ with a width of 0.5 mag and measured each star's color residual relative to the median of its bin. After $3\,\rm{NMAD}$ clipping, the CMD-width was taken as the NMAD of the combined residuals from all valid bins. The reduction in CMD-width before and after dereddening is denoted by $\Delta E_{\rm CMD}$. 
Among the 435 clusters with reliable measurements, 369 (85\%) show a decrease in CMD-width. As shown in Figure~\ref{fig:7}, clusters with stronger differential reddening tend to exhibit larger CMD-width reductions. 
We also tested whether $\Delta E_{\rm CMD}$ depends on cluster angular size ($r_{\rm 50}$) or age, since both may be related to differential reddening. No single threshold in either parameter reliably predicts the amount of CMD-width reduction, indicating that the CMD response depends on the detailed reddening pattern rather than on cluster size or age alone.

The interpretation of any age dependence in the differential reddening amplitude is especially complicated for young OCs.
In this work, differential reddening is measured from the star-to-star scatter in the individual color excesses of cluster members and represents the total reddening variation along the sight lines to the member stars. For older OCs, whose natal gas and dust have largely dispersed, this variation is expected to be dominated by foreground dust inhomogeneities along the sight line. In contrast, young OCs may still be associated with natal molecular clouds or residual local dust, so their measured differential reddening can include both foreground and cluster-associated material. We also checked the dependence of $\mathrm{MAD}(E_{\rm BP-RP})$ on cluster age. The youngest clusters tend to show larger differential reddening than older clusters, consistent with their association with dust-rich environments. However, this trend should not be interpreted as direct evidence for embedded dust alone, because young clusters are also preferentially located close to the Galactic plane, where the foreground dust column and small-scale dust inhomogeneity are larger. We therefore do not attempt to decompose the measured differential reddening into foreground and cluster-associated components.

A potential concern for very young clusters is that circumstellar material around individual members may cause additional reddening and artificially broaden the measured differential reddening. 
We examined clusters containing at least one identified infrared-excess candidate. Using the 2MASS--AllWISE color criteria of \citet{2014ApJ...791..131K}, we identified possible circumstellar disc or envelope sources among members with reliable infrared photometry and repeated the analysis after excluding them. These 26 clusters span $\log(\rm{age/yr})=6.5$--$7.8$ and are therefore predominantly young. The changes in $E(G_{\rm BP}-G_{\rm RP})$ and $\mathrm{MAD}(E_{\rm BP-RP})$ are both close to zero, indicating that these candidates do not significantly affect our differential-reddening estimates. By contrast, the changes in $E(G_{\rm BP}-K_{\rm S})$, $\mathrm{MAD}(E_{\rm BP-K_{\rm S}})$, and the CER are $-0.030\pm0.045$ mag, $-0.006\pm0.023$ mag, and $-0.052\pm0.047$, respectively. The negative offsets suggest that infrared excess tends to increase $E(G_{\rm BP}-K_{\rm S})$ and the CER, likely because near-infrared excess emission from circumstellar material can brighten the $K_{\rm S}$ band.

\subsection{Color Excess Ratios and Extinction Law}\label{sec:4.4}
To characterize the extinction law with reliable cluster CER measurements, we apply additional quality cuts to the color excess sample defined in Section \ref{sec:4.1}. Specifically, we require the CER uncertainty to be smaller than 0.2, $SNR >10$, and $R^2>0.9$. After these cuts, 709 OCs remain in the CER subsample. This selection reduces the contribution of low-quality CER measurements to the statistical properties and spatial trends of $k_{\rm oc}$. We also restrict the sample to $|b|\leq10^\circ$ to focus on sight lines dominated by dust in the Galactic disk, maintain sufficient reddening leverage for CER measurements, and reduce the mixing of dust environments at different vertical heights, leaving 600 OCs for the spatial analysis.

\begin{figure}[!htbp]
\centering
\includegraphics[scale=0.7]{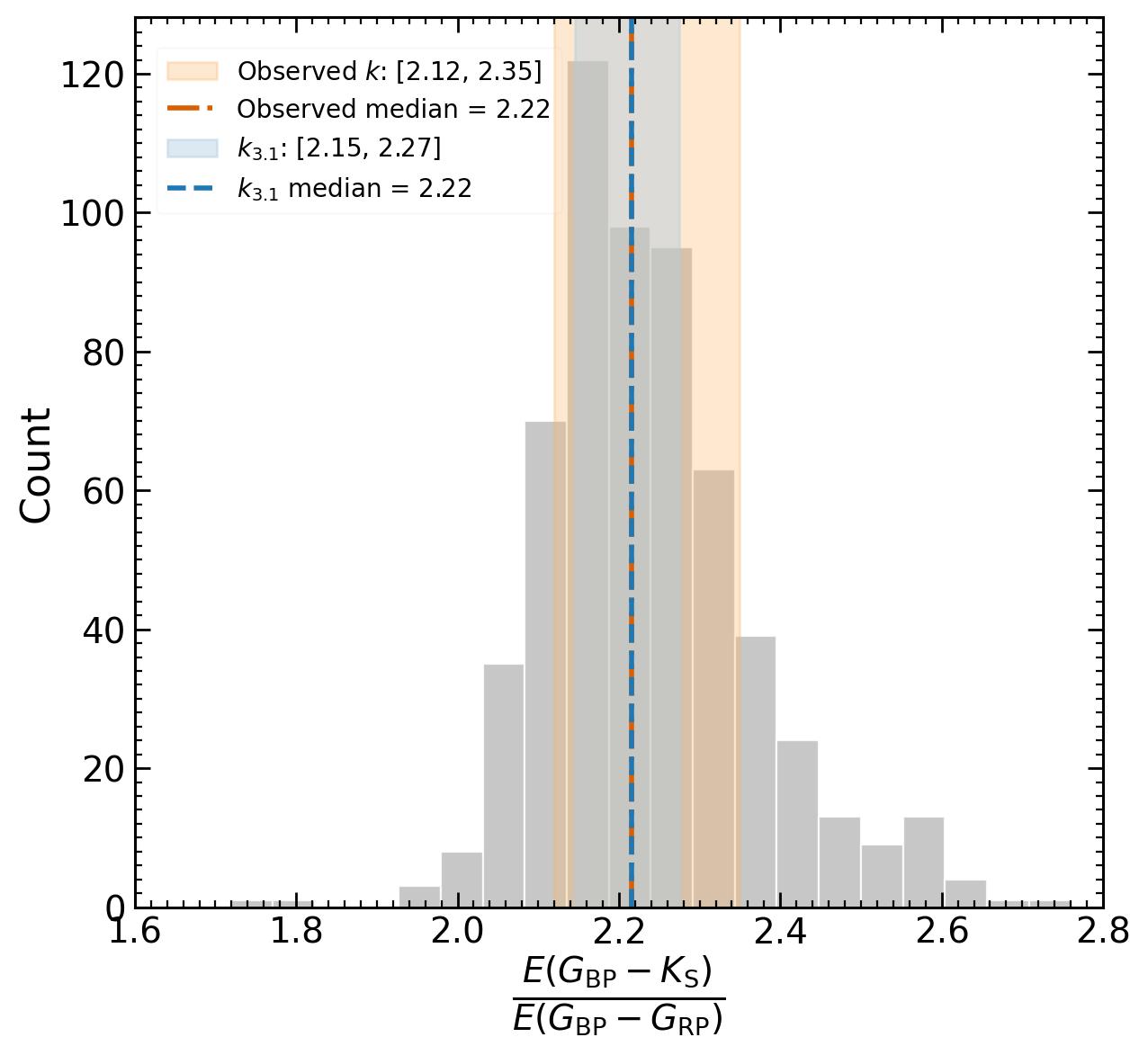}
\figurenum{8}
\caption{Distribution of the cluster CER for the quality-selected OC sample. The orange dashed line denotes the median of the observed cluster CER, and the orange shaded region marks its $P_{16}-P_{84}$ interval. The blue dashed line and shaded region denote, respectively, the median and $P_{16}-P_{84}$ interval of the CER converted from a fiducial $R_{\rm V}=3.1$ extinction law. The nearby $R_{\rm V}=2.9$--3.2 cases produce overlapping CER ranges and are discussed in the text.
}
\label{fig:8}
\end{figure}

We first present the observed CER distribution and compare its scale with standard extinction curve expectations. Figure \ref{fig:8} shows the distribution of the cluster CER $k_{\rm oc}$ for the quality-selected sample. The distribution has a median value of 2.22, with the 16th and 84th percentiles of 2.12 and 2.35, respectively. For comparison, we convert a fiducial diffuse ISM extinction law with $R_{\rm V}=3.1$ into the corresponding CER using the bandpass calculation described below. The resulting 16th--84th percentile interval is $k_{3.1}=2.15$--$2.27$, with a median value of 2.22, matching the observed CER of $2.22^{+0.13}_{-0.10}$. We also repeated the calculation for extinction curves with $R_{\rm V}=2.9$--$3.2$, obtaining median CER values of 2.16--2.24 that vary systematically with $R_{\rm V}$, while their 16th--84th percentile ranges overlap. 
Thus, the CER scale is consistent with a standard $R_{\rm V}=3.1$ extinction curve, but the mean CER alone cannot uniquely determine $R_{\rm V}$ or distinguish it from nearby extinction curves. We therefore discuss the spatial behavior in terms of the observed CER rather than converting individual CER measurements into local $R_{\rm V}$ values. The observed CER distribution provides the reference for examining spatial variations across the Galactic disk.

For this reference CER scale, we compute the bandpass extinction in band $X$ as
\begin{equation}
A_X=-2.5\log_{10}\left[\frac{\int F_\lambda(\lambda)\,S_X(\lambda)\,\lambda\,10^{-0.4A(\lambda)}\,d\lambda}
{\int F_\lambda(\lambda)\,S_X(\lambda)\,\lambda\,d\lambda}\right],
\end{equation}
where $F_\lambda (\lambda)$ is the intrinsic stellar flux and $S_X(\lambda)$ is the filter transmission curve. We adopt \textit{Gaia} DR3 $G_{\rm BP}$, $G_{\rm RP}$ and \textit{2MASS} $K_{\rm S}$ passbands from the SVO Filter Profile Service, and use PHOENIX model atmospheres \citep{2013A&A...553A...6H} to represent the intrinsic SED, sampling a grid of $T_{\rm eff}=4000$--$8000$\,K, $\log g=1$--$5$, $[{\rm M/H}]=(-0.2,0.0,0.3)$ dex, and $A_{\rm V}=0.2$--$4.0$\,mag to account for SED and extinction dependencies.

\begin{figure}[!htbp]
\centering
\includegraphics[width=0.8\textwidth]{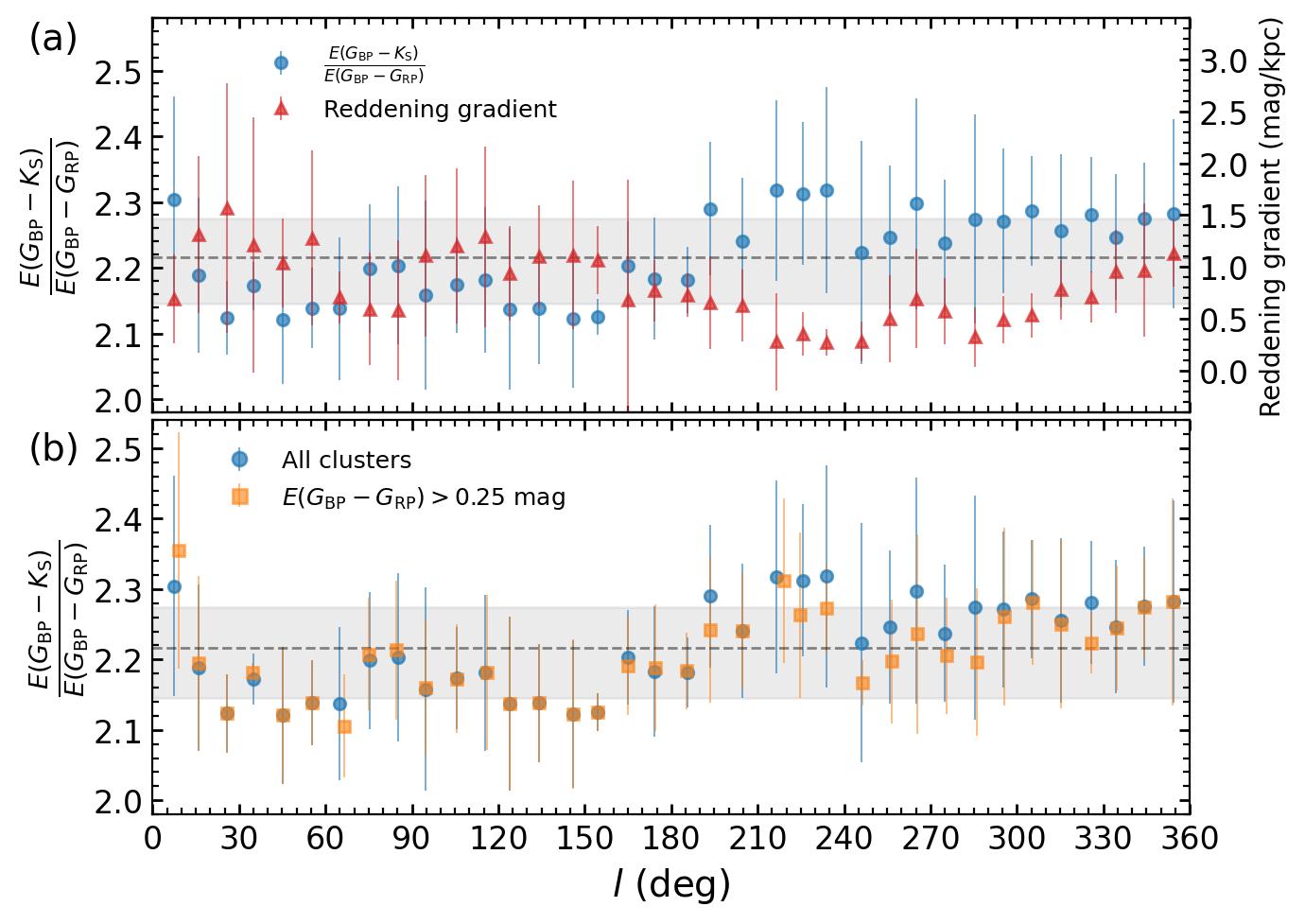}
\figurenum{9}
\caption{
Variation of the OC CER and reddening gradient with Galactic longitude $l$.
Panel (a) shows the binned CER (left axis) and the reddening gradient $E(G_{\rm BP}-K_{\rm S})/d$ (right axis) as a function of $l$, where both quantities are computed as the median within $10^\circ$ bins. 
Panel (b) compares the binned CER for the full near-plane sample (blue circles) with that of the higher-reddening subsample with $E(G_{\rm BP}-G_{\rm RP}) > 0.25$ mag (orange squares). 
Error bars represent the standard deviation within each bin.
The horizontal dashed line and shaded region indicate the reference CER value and its uncertainty range.}
\label{fig:9}
\end{figure}

Figure~\ref{fig:1}(b) shows that this CER distribution is not spatially uniform. Clusters in the first and second Galactic quadrants tend to exhibit lower CER values than those in the third and fourth quadrants. The $X$--$Y$ map shows this large-scale spatial pattern, but the cluster CER is integrated along the line of sight up to the cluster distance. We therefore examine its statistical behavior as a function of Galactic longitude. We divide the sample into longitude bins of $10^\circ$ and compute the median and percentile range of $k_{\rm oc}$ in each bin. As shown in Figure \ref{fig:9}, $k_{\rm oc}$ exhibits a systematic variation with $l$, with smaller values in the first--second Galactic quadrants and larger values in the third--fourth quadrants. This behavior is consistent with the spatial pattern seen in the $X$--$Y$ map, indicating that the azimuthal variation of CER reflects large-scale differences in the dust sampled across the Galactic plane.

Superimposed on this large-scale trend, $k_{\rm oc}$ also shows local fluctuations with longitude. As shown in Figure~\ref{fig:9}(a), directions with larger reddening gradients tend to show smaller $k_{\rm oc}$, while directions with lower reddening gradients more often show larger values. This behavior is consistent with the definition of CER, which depends on the relative scaling of color excesses and is sensitive to the underlying reddening conditions. The high reddening subsample ($E(G_{\rm BP}-G_{\rm RP})_{\rm oc} > 0.25$ mag) supports this interpretation: when low reddening clusters are excluded, the amplitude of the longitude-dependent variation becomes weaker (Figure~\ref{fig:9}(b)), indicating that part of the local variation in $k_{\rm oc}(l)$ is affected by the reddening distribution of the sample.

The large-scale quadrant dependence is broadly consistent with the $R_{\rm V}$ distribution inferred from the dust map of \citet{2025Sci...387.1209Z}, which likewise shows generally lower values in the first and second quadrants and higher values in the third and fourth quadrants. We therefore interpret the spatial pattern of CER as a statistical feature of the OC sample near the Galactic plane, while avoiding a direct conversion of individual CER values into local $R_{\rm V}$ estimates.

The observed quadrant dependence and longitude-dependent CER variations suggest that the optical-to-near-infrared extinction behavior is not spatially uniform across the Galactic disk. Because cluster CERs are integrated along the line of sight up to the cluster distances, these variations should not be interpreted as purely local extinction laws at the cluster positions. Instead, they most likely reflect differences in the dominant dust environments sampled along different Galactic directions. \citet{2026ApJ..1001..114W} showed that the \textit{Gaia} G-band extinction coefficient of Galactic Cepheids varies with reddening, likely due to both nonlinear effects in the broad \textit{Gaia} bands and variations in $R_{\rm V}$ across different interstellar environments. They also showed that such variations can affect optical distance determinations, whereas infrared-based distances are less sensitive to these effects. Our results support accounting for extinction law variations and using infrared observations in studies of the precision distance scale.

\section{Summary}\label{sec:summary}
We present a homogeneous, member-based characterization of extinction properties in Galactic OCs, including the mean reddening, differential reddening, and CER. Using \textit{Gaia}-era open cluster member samples together with \textit{Gaia} DR3 and \textit{2MASS} photometry and intrinsic colors estimated from SHBoost stellar parameters, we obtain cluster-scale extinction measurements under a uniform analysis framework. 
The resulting catalog reports statistical and systematic uncertainties separately, with the systematic uncertainties of the reddening and CER quantities estimated from tests of the adopted stellar-parameter scale. Our main results are as follows. 

\begin{enumerate}
\item For 729 OCs with color excess measurements from more than ten member stars, the cluster reddening spans a broad range across the Galactic plane, from nearly zero to above 2 mag in $E(G_{\rm BP}-G_{\rm RP})$, and is broadly consistent with independent extinction measurements, including literature $E(G_{\rm BP}-G_{\rm RP})$ values for OCs \citep{2025AJ....169..115W}, $A_{\rm V}$ values for OCs \citep{2020A&A...640A...1C} and the three-dimensional dust map of \citet{2025Sci...387.1209Z}. This agreement supports the robustness of our reddening measurements, while the member-based framework preserves cluster-scale reddening structure that is largely smoothed out in coarse dust maps and cannot be recovered from global CMD fitting.

\item The individual reddening estimates improve CMD-based analyses of clusters affected by strong differential reddening. Star-by-star corrections recover narrower and more coherent dereddened CMD sequences, while cluster color excess maps reveal small-scale extinction structures that are only weakly represented in lower-resolution dust maps. The increase of differential reddening with mean reddening indicates that sight lines with larger dust columns tend to show stronger small-scale extinction inhomogeneity. 

\item For the quality-selected CER subsample of 600 OCs near the Galactic plane, CER exhibits clear large-scale spatial variations across the Galactic disk, with systematically lower values in the first and second Galactic quadrants than in the third and fourth quadrants. Because CER is integrated along the line of sight, these variations are best interpreted as reflecting differences in the dominant dust environments sampled along different directions, rather than local extinction properties at individual cluster positions.
These spatial variations indicate that the optical-to-near-infrared extinction behavior is not uniform across the Galactic disk, with implications for studies of the extinction law and precision distance measurements.
\end{enumerate} 

Future studies incorporating additional photometric bands and improved distance-resolved analyses will provide stronger constraints on the large-scale distribution of interstellar dust and extinction law variations across the Galactic disk, while enabling more accurate and consistent OC parameter determination. 

\begin{acknowledgments} 
We thank the anonymous referee for the helpful comments. This work is supported by the National Natural Science Foundation of China (NSFC) through the projects 12373028, 12322306, 12173047, and 12133002. This work is also supported by the science research grants from the China Manned Space Project with No. CMS-CSST-2025-A01. S.W. and X.C. acknowledge support from the Youth Innovation Promotion Association of the CAS (grant Nos. 2023065 and 2022055). This work has made use of data from the European Space Agency (ESA) mission \textit{Gaia} (\url{https://www.cosmos.esa.int/gaia}),
processed by the \textit{Gaia} Data Processing and Analysis Consortium (DPAC, \url{https://www.cosmos.esa.int/web/gaia/dpac/consortium}).
Funding for the DPAC has been provided by national institutions, in particular the institutions participating in the \textit{Gaia} Multilateral Agreement.

\end{acknowledgments}

\bibliography{OC}{}

@ARTICLE{2023ApJS..267....8A,
       author = {{Andrae}, Ren{\'e} and {Rix}, Hans-Walter and {Chandra}, Vedant},
        title = "{Robust Data-driven Metallicities for 175 Million Stars from Gaia XP Spectra}",
      journal = {\apjs},
         year = 2023,
        month = jul,
       volume = {267},
       number = {1},
          eid = {8},
        pages = {8},
          doi = {10.3847/1538-4365/acd53e},
archivePrefix = {arXiv},
       eprint = {2302.02611},
 primaryClass = {astro-ph.SR},
       adsurl = {https://ui.adsabs.harvard.edu/abs/2023ApJS..267....8A}
}

@ARTICLE{2023MNRAS.524.1855Z,
       author = {{Zhang}, Xiangyu and {Green}, Gregory M. and {Rix}, Hans-Walter},
        title = "{Parameters of 220 million stars from Gaia BP/RP spectra}",
      journal = {\mnras},
         year = 2023,
        month = sep,
       volume = {524},
       number = {2},
        pages = {1855-1884},
          doi = {10.1093/mnras/stad1941},
archivePrefix = {arXiv},
       eprint = {2303.03420},
 primaryClass = {astro-ph.SR},
       adsurl = {https://ui.adsabs.harvard.edu/abs/2023MNRAS.524.1855Z}
}

@ARTICLE{2024MNRAS.52710937Y,
       author = {{Yao}, Yupeng and {Ji}, Alexander P. and {Koposov}, Sergey E. and {Limberg}, Guilherme},
        title = "{200 000 candidate very metal-poor stars in Gaia DR3 XP spectra}",
      journal = {\mnras},
         year = 2024,
        month = feb,
       volume = {527},
       number = {4},
        pages = {10937-10954},
          doi = {10.1093/mnras/stad3775},
archivePrefix = {arXiv},
       eprint = {2303.17676},
 primaryClass = {astro-ph.GA},
       adsurl = {https://ui.adsabs.harvard.edu/abs/2024MNRAS.52710937Y}
}

@ARTICLE{2020AA...633A..99C,
       author = {{Cantat-Gaudin}, T. and {Anders}, F.},
        title = "{Clusters and mirages: cataloguing stellar aggregates in the Milky Way}",
      journal = {\aap},
         year = 2020,
        month = jan,
       volume = {633},
          eid = {A99},
        pages = {A99},
          doi = {10.1051/0004-6361/201936691},
archivePrefix = {arXiv},
       eprint = {1911.07075},
 primaryClass = {astro-ph.SR},
       adsurl = {https://ui.adsabs.harvard.edu/abs/2020A&A...633A..99C}
}

@ARTICLE{2020A&A...640A...1C,
       author = {{Cantat-Gaudin}, T. and {Anders}, F. and {Castro-Ginard}, A. and {Jordi}, C. and {Romero-G{\'o}mez}, M. and {Soubiran}, C. and {Casamiquela}, L. and {Tarricq}, Y. and {Moitinho}, A. and {Vallenari}, A. and {Bragaglia}, A. and {Krone-Martins}, A. and {Kounkel}, M.},
        title = "{Painting a portrait of the Galactic disc with its stellar clusters}",
      journal = {\aap},
         year = 2020,
        month = aug,
       volume = {640},
          eid = {A1},
        pages = {A1},
          doi = {10.1051/0004-6361/202038192},
archivePrefix = {arXiv},
       eprint = {2004.07274},
 primaryClass = {astro-ph.GA},
       adsurl = {https://ui.adsabs.harvard.edu/abs/2020A&A...640A...1C}
}

@ARTICLE{2025Sci...387.1209Z,
       author = {{Zhang}, Xiangyu and {Green}, Gregory M.},
        title = "{Three-dimensional maps of the interstellar dust extinction curve within the Milky Way galaxy}",
      journal = {Science},
         year = 2025,
        month = mar,
       volume = {387},
       number = {6739},
        pages = {1209-1214},
          doi = {10.1126/science.ado9787},
archivePrefix = {arXiv},
       eprint = {2407.14594},
 primaryClass = {astro-ph.GA},
       adsurl = {https://ui.adsabs.harvard.edu/abs/2025Sci...387.1209Z}
}

@ARTICLE{2014ApJ...783..130R,
       author = {{Reid}, M.~J. and {Menten}, K.~M. and {Brunthaler}, A. and {Zheng}, X.~W. and {Dame}, T.~M. and {Xu}, Y. and {Wu}, Y. and {Zhang}, B. and {Sanna}, A. and {Sato}, M. and {Hachisuka}, K. and {Choi}, Y.~K. and {Immer}, K. and {Moscadelli}, L. and {Rygl}, K.~L.~J. and {Bartkiewicz}, A.},
        title = "{Trigonometric Parallaxes of High Mass Star Forming Regions: The Structure and Kinematics of the Milky Way}",
      journal = {\apj},
         year = 2014,
        month = mar,
       volume = {783},
       number = {2},
          eid = {130},
        pages = {130},
          doi = {10.1088/0004-637X/783/2/130},
archivePrefix = {arXiv},
       eprint = {1401.5377},
 primaryClass = {astro-ph.GA},
       adsurl = {https://ui.adsabs.harvard.edu/abs/2014ApJ...783..130R}
}

@ARTICLE{2024A&A...691A..98K,
       author = {{Khalatyan}, A. and {Anders}, F. and {Chiappini}, C. and {Queiroz}, A.~B.~A. and {Nepal}, S. and {dal Ponte}, M. and {Jordi}, C. and {Guiglion}, G. and {Valentini}, M. and {Torralba Elipe}, G. and {Steinmetz}, M. and {Pantaleoni-Gonz{\'a}lez}, M. and {Malhotra}, S. and {Jim{\'e}nez-Arranz}, {\'O}. and {Enke}, H. and {Casamiquela}, L. and {Ard{\`e}vol}, J.},
        title = "{Transferring spectroscopic stellar labels to 217 million Gaia DR3 XP stars with SHBoost}",
      journal = {\aap},
         year = 2024,
        month = nov,
       volume = {691},
          eid = {A98},
        pages = {A98},
          doi = {10.1051/0004-6361/202451427},
archivePrefix = {arXiv},
       eprint = {2407.06963},
 primaryClass = {astro-ph.SR},
       adsurl = {https://ui.adsabs.harvard.edu/abs/2024A&A...691A..98K}
}

@ARTICLE{2014ApJ...788L..12W,
       author = {{Wang}, Shu and {Jiang}, B.~W.},
        title = "{Universality of the Near-infrared Extinction Law Based on the APOGEE Survey}",
      journal = {\apjl},
         year = 2014,
        month = jun,
       volume = {788},
       number = {1},
          eid = {L12},
        pages = {L12},
          doi = {10.1088/2041-8205/788/1/L12},
       adsurl = {https://ui.adsabs.harvard.edu/abs/2014ApJ...788L..12W}
}

@ARTICLE{2019ApJ...877..116W,
       author = {{Wang}, Shu and {Chen}, Xiaodian},
        title = "{The Optical to Mid-infrared Extinction Law Based on the APOGEE, Gaia DR2, Pan-STARRS1, SDSS, APASS, 2MASS, and WISE Surveys}",
      journal = {\apj},
         year = 2019,
        month = jun,
       volume = {877},
       number = {2},
          eid = {116},
        pages = {116},
          doi = {10.3847/1538-4357/ab1c61},
archivePrefix = {arXiv},
       eprint = {1904.04575},
 primaryClass = {astro-ph.GA},
       adsurl = {https://ui.adsabs.harvard.edu/abs/2019ApJ...877..116W}
}

@ARTICLE{2023ApJ...956...26L,
       author = {{Li}, Ling and {Wang}, Shu and {Chen}, Xiaodian and {Jiang}, QingQuan},
        title = "{The Ultraviolet to Mid-infrared Extinction Law of the Taurus Molecular Cloud Based on the Gaia DR3, GALEX, APASS, Pan-STARRS1, 2MASS, and WISE Surveys}",
      journal = {\apj},
         year = 2023,
        month = oct,
       volume = {956},
       number = {1},
          eid = {26},
        pages = {26},
          doi = {10.3847/1538-4357/aced8a},
archivePrefix = {arXiv},
       eprint = {2308.02156},
 primaryClass = {astro-ph.GA},
       adsurl = {https://ui.adsabs.harvard.edu/abs/2023ApJ...956...26L}
}

@ARTICLE{2023ApJ...946...43W,
       author = {{Wang}, Shu and {Chen}, Xiaodian},
        title = "{The Optical to Infrared Extinction Law of Magellanic Clouds Based on Red Supergiants and Classical Cepheids}",
      journal = {\apj},
         year = 2023,
        month = mar,
       volume = {946},
       number = {1},
          eid = {43},
        pages = {43},
          doi = {10.3847/1538-4357/acb647},
archivePrefix = {arXiv},
       eprint = {2301.09146},
 primaryClass = {astro-ph.GA},
       adsurl = {https://ui.adsabs.harvard.edu/abs/2023ApJ...946...43W}
}

@ARTICLE{2025ApJ...982...77D,
       author = {{Deng}, Juan and {Wang}, Shu and {Jiang}, Biwei and {Zhao}, He},
        title = "{The Multiwavelength Extinction Law and Its Variation in the Coalsack Molecular Cloud Based on the Gaia, APASS, SMSS, 2MASS, GLIMPSE, and WISE Surveys}",
      journal = {\apj},
         year = 2025,
        month = apr,
       volume = {982},
       number = {2},
          eid = {77},
        pages = {77},
          doi = {10.3847/1538-4357/adb431},
archivePrefix = {arXiv},
       eprint = {2502.08956},
 primaryClass = {astro-ph.GA},
       adsurl = {https://ui.adsabs.harvard.edu/abs/2025ApJ...982...77D}
}

@ARTICLE{2024ApJ...974..138Z,
       author = {{Zhao}, He and {Wang}, Shu and {Jiang}, Biwei and {Li}, Jun and {Fan}, Dongwei and {Ren}, Yi and {Ma}, Xiaoxiao},
        title = "{Data-driven Stellar Intrinsic Colors and Dust Reddenings for Spectrophotometric Data: From the Blue-edge Method to a Machine Learning Approach}",
      journal = {\apj},
         year = 2024,
        month = oct,
       volume = {974},
       number = {1},
          eid = {138},
        pages = {138},
          doi = {10.3847/1538-4357/ad6d64},
archivePrefix = {arXiv},
       eprint = {2407.17386},
 primaryClass = {astro-ph.SR},
       adsurl = {https://ui.adsabs.harvard.edu/abs/2024ApJ...974..138Z}
}

@ARTICLE{2023A&A...673A.114H,
       author = {{Hunt}, Emily L. and {Reffert}, Sabine},
        title = "{Improving the open cluster census. II. An all-sky cluster catalogue with Gaia DR3}",
      journal = {\aap},
         year = 2023,
        month = may,
       volume = {673},
          eid = {A114},
        pages = {A114},
          doi = {10.1051/0004-6361/202346285},
archivePrefix = {arXiv},
       eprint = {2303.13424},
 primaryClass = {astro-ph.GA},
       adsurl = {https://ui.adsabs.harvard.edu/abs/2023A&A...673A.114H}
}

@ARTICLE{2024A&A...686A..42H,
       author = {{Hunt}, Emily L. and {Reffert}, Sabine},
        title = "{Improving the open cluster census. III. Using cluster masses, radii, and dynamics to create a cleaned open cluster catalogue}",
      journal = {\aap},
         year = 2024,
        month = jun,
       volume = {686},
          eid = {A42},
        pages = {A42},
          doi = {10.1051/0004-6361/202348662},
archivePrefix = {arXiv},
       eprint = {2403.05143},
 primaryClass = {astro-ph.GA},
       adsurl = {https://ui.adsabs.harvard.edu/abs/2024A&A...686A..42H}
}

@ARTICLE{2025AJ....170..288L,
       author = {{Li}, Lu and {Shao}, Zhengyi and {Li}, Zhaozhou and {Fu}, Xiaoting},
        title = "{The MiMO Catalog: Physical Parameters and Stellar Mass Functions of 1232 Open Clusters from Gaia DR3}",
      journal = {\aj},
         year = 2025,
        month = nov,
       volume = {170},
       number = {5},
          eid = {288},
        pages = {288},
          doi = {10.3847/1538-3881/ae0cb6},
archivePrefix = {arXiv},
       eprint = {2510.23374},
 primaryClass = {astro-ph.GA},
       adsurl = {https://ui.adsabs.harvard.edu/abs/2025AJ....170..288L}
}

@ARTICLE{2025A&A...703A.100N,
       author = {{Nizovkina}, M. and {Larsen}, S.~S. and {Brown}, A.~G.~A. and {Helmi}, A.},
        title = "{Refining open cluster parameters with Gaia XP metallicities}",
      journal = {\aap},
         year = 2025,
        month = nov,
       volume = {703},
          eid = {A100},
        pages = {A100},
          doi = {10.1051/0004-6361/202555437},
archivePrefix = {arXiv},
       eprint = {2510.10385},
 primaryClass = {astro-ph.GA},
       adsurl = {https://ui.adsabs.harvard.edu/abs/2025A&A...703A.100N}
}

@ARTICLE{2024AJ....167...12C,
       author = {{Cavallo}, Lorenzo and {Spina}, Lorenzo and {Carraro}, Giovanni and {Magrini}, Laura and {Poggio}, Eloisa and {Cantat-Gaudin}, Tristan and {Pasquato}, Mario and {Lucatello}, Sara and {Ortolani}, Sergio and {Schiappacasse-Ulloa}, Jose},
        title = "{Parameter Estimation for Open Clusters using an Artificial Neural Network with a QuadTree-based Feature Extractor}",
      journal = {\aj},
         year = 2024,
        month = jan,
       volume = {167},
       number = {1},
          eid = {12},
        pages = {12},
          doi = {10.3847/1538-3881/ad07e5},
archivePrefix = {arXiv},
       eprint = {2311.03009},
 primaryClass = {astro-ph.GA},
       adsurl = {https://ui.adsabs.harvard.edu/abs/2024AJ....167...12C}
}

@ARTICLE{2019AJ....158..122K,
       author = {{Kounkel}, Marina and {Covey}, Kevin},
        title = "{Untangling the Galaxy. I. Local Structure and Star Formation History of the Milky Way}",
      journal = {\aj},
         year = 2019,
        month = sep,
       volume = {158},
       number = {3},
          eid = {122},
        pages = {122},
          doi = {10.3847/1538-3881/ab339a},
archivePrefix = {arXiv},
       eprint = {1907.07709},
 primaryClass = {astro-ph.GA},
       adsurl = {https://ui.adsabs.harvard.edu/abs/2019AJ....158..122K}
}

@ARTICLE{2017PASA...34...68R,
       author = {{Rangwal}, Geeta and {Yadav}, R.~K.~S. and {Durgapal}, Alok K. and {Bisht}, D.},
        title = "{Interstellar Extinction in 20 Open Star Clusters}",
      journal = {\pasa},
         year = 2017,
        month = dec,
       volume = {34},
          eid = {e068},
        pages = {e068},
          doi = {10.1017/pasa.2017.64},
archivePrefix = {arXiv},
       eprint = {1711.09591},
 primaryClass = {astro-ph.GA},
       adsurl = {https://ui.adsabs.harvard.edu/abs/2017PASA...34...68R}
}

@ARTICLE{2003A&A...397..191P,
       author = {{Pandey}, A.~K. and {Upadhyay}, K. and {Nakada}, Y. and {Ogura}, K.},
        title = "{Interstellar extinction in the open clusters towards galactic longitude around 130$^{deg}$}",
      journal = {\aap},
         year = 2003,
        month = jan,
       volume = {397},
        pages = {191-200},
          doi = {10.1051/0004-6361:20021509},
archivePrefix = {arXiv},
       eprint = {astro-ph/0210680},
 primaryClass = {astro-ph},
       adsurl = {https://ui.adsabs.harvard.edu/abs/2003A&A...397..191P}
}

@ARTICLE{2024NewAR..9901696C,
       author = {{Cantat-Gaudin}, T. and {Casamiquela}, L.},
        title = "{How Gaia sheds light on the Milky Way star cluster population}",
      journal = {\nar},
         year = 2024,
        month = dec,
       volume = {99},
          eid = {101696},
        pages = {101696},
          doi = {10.1016/j.newar.2024.101696},
archivePrefix = {arXiv},
       eprint = {2406.03308},
 primaryClass = {astro-ph.GA},
       adsurl = {https://ui.adsabs.harvard.edu/abs/2024NewAR..9901696C}
}

@ARTICLE{2009ApJ...707..510Z,
       author = {{Zasowski}, G. and {Majewski}, S.~R. and {Indebetouw}, R. and {Meade}, M.~R. and {Nidever}, D.~L. and {Patterson}, R.~J. and {Babler}, B. and {Skrutskie}, M.~F. and {Watson}, C. and {Whitney}, B.~A. and {Churchwell}, E.},
        title = "{Lifting the Dusty Veil with Near- and Mid-Infrared Photometry. II. A Large-Scale Study of the Galactic Infrared Extinction Law}",
      journal = {\apj},
         year = 2009,
        month = dec,
       volume = {707},
       number = {1},
        pages = {510-523},
          doi = {10.1088/0004-637X/707/1/510},
archivePrefix = {arXiv},
       eprint = {0910.4403},
 primaryClass = {astro-ph.GA},
       adsurl = {https://ui.adsabs.harvard.edu/abs/2009ApJ...707..510Z}
}

@ARTICLE{2014A&A...561A..57K,
       author = {{Krone-Martins}, A. and {Moitinho}, A.},
        title = "{UPMASK: unsupervised photometric membership assignment in stellar clusters}",
      journal = {\aap},
         year = 2014,
        month = jan,
       volume = {561},
          eid = {A57},
        pages = {A57},
          doi = {10.1051/0004-6361/201321143},
archivePrefix = {arXiv},
       eprint = {1309.4471},
 primaryClass = {astro-ph.IM},
       adsurl = {https://ui.adsabs.harvard.edu/abs/2014A&A...561A..57K}
}

@ARTICLE{2018A&A...618A..93C,
       author = {{Cantat-Gaudin}, T. and {Jordi}, C. and {Vallenari}, A. and {Bragaglia}, A. and {Balaguer-N{\'u}{\~n}ez}, L. and {Soubiran}, C. and {Bossini}, D. and {Moitinho}, A. and {Castro-Ginard}, A. and {Krone-Martins}, A. and {Casamiquela}, L. and {Sordo}, R. and {Carrera}, R.},
        title = "{A Gaia DR2 view of the open cluster population in the Milky Way}",
      journal = {\aap},
         year = 2018,
        month = oct,
       volume = {618},
          eid = {A93},
        pages = {A93},
          doi = {10.1051/0004-6361/201833476},
archivePrefix = {arXiv},
       eprint = {1805.08726},
 primaryClass = {astro-ph.GA},
       adsurl = {https://ui.adsabs.harvard.edu/abs/2018A&A...618A..93C}
}

@ARTICLE{2018SSRv..214...74M,
       author = {{Matsunaga}, Noriyuki and {Bono}, Giuseppe and {Chen}, Xiaodian and {de Grijs}, Richard and {Inno}, Laura and {Nishiyama}, Shogo},
        title = "{Impact of Distance Determinations on Galactic Structure. I. Young and Intermediate-Age Tracers}",
      journal = {\ssr},
         year = 2018,
        month = jun,
       volume = {214},
       number = {4},
          eid = {74},
        pages = {74},
          doi = {10.1007/s11214-018-0506-5},
archivePrefix = {arXiv},
       eprint = {1804.04931},
 primaryClass = {astro-ph.SR},
       adsurl = {https://ui.adsabs.harvard.edu/abs/2018SSRv..214...74M}
}

@ARTICLE{2016ApJ...821...78S,
       author = {{Schlafly}, E.~F. and {Meisner}, A.~M. and {Stutz}, A.~M. and {Kainulainen}, J. and {Peek}, J.~E.~G. and {Tchernyshyov}, K. and {Rix}, H.-W. and {Finkbeiner}, D.~P. and {Covey}, K.~R. and {Green}, G.~M. and {Bell}, E.~F. and {Burgett}, W.~S. and {Chambers}, K.~C. and {Draper}, P.~W. and {Flewelling}, H. and {Hodapp}, K.~W. and {Kaiser}, N. and {Magnier}, E.~A. and {Martin}, N.~F. and {Metcalfe}, N. and {Wainscoat}, R.~J. and {Waters}, C.},
        title = "{The Optical-infrared Extinction Curve and Its Variation in the Milky Way}",
      journal = {\apj},
         year = 2016,
        month = apr,
       volume = {821},
       number = {2},
          eid = {78},
        pages = {78},
          doi = {10.3847/0004-637X/821/2/78},
archivePrefix = {arXiv},
       eprint = {1602.03928},
 primaryClass = {astro-ph.GA},
       adsurl = {https://ui.adsabs.harvard.edu/abs/2016ApJ...821...78S}
}

@ARTICLE{2006AJ....131.1163S,
       author = {{Skrutskie}, M.~F. and {Cutri}, R.~M. and {Stiening}, R. and {Weinberg}, M.~D. and {Schneider}, S. and {Carpenter}, J.~M. and {Beichman}, C. and {Capps}, R. and {Chester}, T. and {Elias}, J. and {Huchra}, J. and {Liebert}, J. and {Lonsdale}, C. and {Monet}, D.~G. and {Price}, S. and {Seitzer}, P. and {Jarrett}, T. and {Kirkpatrick}, J.~D. and {Gizis}, J.~E. and {Howard}, E. and {Evans}, T. and {Fowler}, J. and {Fullmer}, L. and {Hurt}, R. and {Light}, R. and {Kopan}, E.~L. and {Marsh}, K.~A. and {McCallon}, H.~L. and {Tam}, R. and {Van Dyk}, S. and {Wheelock}, S.},
        title = "{The Two Micron All Sky Survey (2MASS)}",
      journal = {\aj},
         year = 2006,
        month = feb,
       volume = {131},
       number = {2},
        pages = {1163-1183},
          doi = {10.1086/498708},
       adsurl = {https://ui.adsabs.harvard.edu/abs/2006AJ....131.1163S}
}

@ARTICLE{2023A&A...674A...1G,
       author = {{Gaia Collaboration} and {Vallenari}, A. and {Brown}, A.~G.~A. and {Prusti}, T. and {de Bruijne}, J.~H.~J. and {Arenou}, F. and {Babusiaux}, C. and {Biermann}, M. and {Creevey}, O.~L. and {Ducourant}, C. and {Evans}, D.~W. and {Eyer}, L. and {Guerra}, R. and {Hutton}, A. and {Jordi}, C. and {Klioner}, S.~A. and {Lammers}, U.~L. and {Lindegren}, L. and {Luri}, X. and {Mignard}, F. and {Panem}, C. and {Pourbaix}, D. and {Randich}, S. and {Sartoretti}, P. and {Soubiran}, C. and {Tanga}, P. and {Walton}, N.~A. and {Bailer-Jones}, C.~A.~L. and {Bastian}, U. and {Drimmel}, R. and {Jansen}, F. and {Katz}, D. and {Lattanzi}, M.~G. and {van Leeuwen}, F. and {Bakker}, J. and {Cacciari}, C. and {Casta{\~n}eda}, J. and {De Angeli}, F. and {Fabricius}, C. and {Fouesneau}, M. and {Fr{\'e}mat}, Y. and {Galluccio}, L. and {Guerrier}, A. and {Heiter}, U. and {Masana}, E. and {Messineo}, R. and {Mowlavi}, N. and {Nicolas}, C. and {Nienartowicz}, K. and {Pailler}, F. and {Panuzzo}, P. and {Riclet}, F. and {Roux}, W. and {Seabroke}, G.~M. and {Sordo}, R. and {Th{\'e}venin}, F. and {Gracia-Abril}, G. and {Portell}, J. and {Teyssier}, D. and {Altmann}, M. and {Andrae}, R. and {Audard}, M. and {Bellas-Velidis}, I. and {Benson}, K. and {Berthier}, J. and {Blomme}, R. and {Burgess}, P.~W. and {Busonero}, D. and {Busso}, G. and {C{\'a}novas}, H. and {Carry}, B. and {Cellino}, A. and {Cheek}, N. and {Clementini}, G. and {Damerdji}, Y. and {Davidson}, M. and {de Teodoro}, P. and {Nu{\~n}ez Campos}, M. and {Delchambre}, L. and {Dell'Oro}, A. and {Esquej}, P. and {Fern{\'a}ndez-Hern{\'a}ndez}, J. and {Fraile}, E. and {Garabato}, D. and {Garc{\'\i}a-Lario}, P. and {Gosset}, E. and {Haigron}, R. and {Halbwachs}, J.-L. and {Hambly}, N.~C. and {Harrison}, D.~L. and {Hern{\'a}ndez}, J. and {Hestroffer}, D. and {Hodgkin}, S.~T. and {Holl}, B. and {Jan{\ss}en}, K. and {Jevardat de Fombelle}, G. and {Jordan}, S. and {Krone-Martins}, A. and {Lanzafame}, A.~C. and {L{\"o}ffler}, W. and {Marchal}, O. and {Marrese}, P.~M. and {Moitinho}, A. and {Muinonen}, K. and {Osborne}, P. and {Pancino}, E. and {Pauwels}, T. and {Recio-Blanco}, A. and {Reyl{\'e}}, C. and {Riello}, M. and {Rimoldini}, L. and {Roegiers}, T. and {Rybizki}, J. and {Sarro}, L.~M. and {Siopis}, C. and {Smith}, M. and {Sozzetti}, A. and {Utrilla}, E. and {van Leeuwen}, M. and {Abbas}, U. and {{\'A}brah{\'a}m}, P. and {Abreu Aramburu}, A. and {Aerts}, C. and {Aguado}, J.~J. and {Ajaj}, M. and {Aldea-Montero}, F. and {Altavilla}, G. and {{\'A}lvarez}, M.~A. and {Alves}, J. and {Anders}, F. and {Anderson}, R.~I. and {Anglada Varela}, E. and {Antoja}, T. and {Baines}, D. and {Baker}, S.~G. and {Balaguer-N{\'u}{\~n}ez}, L. and {Balbinot}, E. and {Balog}, Z. and {Barache}, C. and {Barbato}, D. and {Barros}, M. and {Barstow}, M.~A. and {Bartolom{\'e}}, S. and {Bassilana}, J.-L. and {Bauchet}, N. and {Becciani}, U. and {Bellazzini}, M. and {Berihuete}, A. and {Bernet}, M. and {Bertone}, S. and {Bianchi}, L. and {Binnenfeld}, A. and {Blanco-Cuaresma}, S. and {Blazere}, A. and {Boch}, T. and {Bombrun}, A. and {Bossini}, D. and {Bouquillon}, S. and {Bragaglia}, A. and {Bramante}, L. and {Breedt}, E. and {Bressan}, A. and {Brouillet}, N. and {Brugaletta}, E. and {Bucciarelli}, B. and {Burlacu}, A. and {Butkevich}, A.~G. and {Buzzi}, R. and {Caffau}, E. and {Cancelliere}, R. and {Cantat-Gaudin}, T. and {Carballo}, R. and {Carlucci}, T. and {Carnerero}, M.~I. and {Carrasco}, J.~M. and {Casamiquela}, L. and {Castellani}, M. and {Castro-Ginard}, A. and {Chaoul}, L. and {Charlot}, P. and {Chemin}, L. and {Chiaramida}, V. and {Chiavassa}, A. and {Chornay}, N. and {Comoretto}, G. and {Contursi}, G. and {Cooper}, W.~J. and {Cornez}, T. and {Cowell}, S. and {Crifo}, F. and {Cropper}, M. and {Crosta}, M. and {Crowley}, C. and {Dafonte}, C. and {Dapergolas}, A. and {David}, M. and {David}, P. and {de Laverny}, P. and {De Luise}, F. and {De March}, R.},
        title = "{Gaia Data Release 3. Summary of the content and survey properties}",
      journal = {\aap},
         year = 2023,
        month = jun,
       volume = {674},
          eid = {A1},
        pages = {A1},
          doi = {10.1051/0004-6361/202243940},
archivePrefix = {arXiv},
       eprint = {2208.00211},
 primaryClass = {astro-ph.GA},
       adsurl = {https://ui.adsabs.harvard.edu/abs/2023A&A...674A...1G}
}

@ARTICLE{2013A&A...553A...6H,
       author = {{Husser}, T.-O. and {Wende-von Berg}, S. and {Dreizler}, S. and {Homeier}, D. and {Reiners}, A. and {Barman}, T. and {Hauschildt}, P.~H.},
        title = "{A new extensive library of PHOENIX stellar atmospheres and synthetic spectra}",
      journal = {\aap},
         year = 2013,
        month = may,
       volume = {553},
          eid = {A6},
        pages = {A6},
          doi = {10.1051/0004-6361/201219058},
archivePrefix = {arXiv},
       eprint = {1303.5632},
 primaryClass = {astro-ph.SR},
       adsurl = {https://ui.adsabs.harvard.edu/abs/2013A&A...553A...6H}
}

@ARTICLE{2021A&A...646A.104H,
       author = {{Hunt}, Emily L. and {Reffert}, Sabine},
        title = "{Improving the open cluster census. I. Comparison of clustering algorithms applied to Gaia DR2 data}",
      journal = {\aap},
         year = 2021,
        month = feb,
       volume = {646},
          eid = {A104},
        pages = {A104},
          doi = {10.1051/0004-6361/202039341},
archivePrefix = {arXiv},
       eprint = {2012.04267},
 primaryClass = {astro-ph.GA},
       adsurl = {https://ui.adsabs.harvard.edu/abs/2021A&A...646A.104H}
}

@ARTICLE{2020A&A...635A..45C,
       author = {{Castro-Ginard}, A. and {Jordi}, C. and {Luri}, X. and {{\'A}lvarez Cid-Fuentes}, J. and {Casamiquela}, L. and {Anders}, F. and {Cantat-Gaudin}, T. and {Mongui{\'o}}, M. and {Balaguer-N{\'u}{\~n}ez}, L. and {Sol{\`a}}, S. and {Badia}, R.~M.},
        title = "{Hunting for open clusters in Gaia DR2: 582 new open clusters in the Galactic disc}",
      journal = {\aap},
         year = 2020,
        month = mar,
       volume = {635},
          eid = {A45},
        pages = {A45},
          doi = {10.1051/0004-6361/201937386},
archivePrefix = {arXiv},
       eprint = {2001.07122},
 primaryClass = {astro-ph.GA},
       adsurl = {https://ui.adsabs.harvard.edu/abs/2020A&A...635A..45C}
}

@ARTICLE{2022A&A...661A.118C,
       author = {{Castro-Ginard}, A. and {Jordi}, C. and {Luri}, X. and {Cantat-Gaudin}, T. and {Carrasco}, J.~M. and {Casamiquela}, L. and {Anders}, F. and {Balaguer-N{\'u}{\~n}ez}, L. and {Badia}, R.~M.},
        title = "{Hunting for open clusters in Gaia EDR3: 628 new open clusters found with OCfinder}",
      journal = {\aap},
         year = 2022,
        month = may,
       volume = {661},
          eid = {A118},
        pages = {A118},
          doi = {10.1051/0004-6361/202142568},
archivePrefix = {arXiv},
       eprint = {2111.01819},
 primaryClass = {astro-ph.GA},
       adsurl = {https://ui.adsabs.harvard.edu/abs/2022A&A...661A.118C}
}

@ARTICLE{2021MNRAS.504..356D,
       author = {{Dias}, W.~S. and {Monteiro}, H. and {Moitinho}, A. and {L{\'e}pine}, J.~R.~D. and {Carraro}, G. and {Paunzen}, E. and {Alessi}, B. and {Villela}, L.},
        title = "{Updated parameters of 1743 open clusters based on Gaia DR2}",
      journal = {\mnras},
         year = 2021,
        month = jun,
       volume = {504},
       number = {1},
        pages = {356-371},
          doi = {10.1093/mnras/stab770},
archivePrefix = {arXiv},
       eprint = {2103.12829},
 primaryClass = {astro-ph.SR},
       adsurl = {https://ui.adsabs.harvard.edu/abs/2021MNRAS.504..356D}
}

@ARTICLE{2021A&A...652A.102H,
       author = {{Hao}, C.~J. and {Xu}, Y. and {Hou}, L.~G. and {Bian}, S.~B. and {Li}, J.~J. and {Wu}, Z.~Y. and {He}, Z.~H. and {Li}, Y.~J. and {Liu}, D.~J.},
        title = "{Evolution of the local spiral structure of the Milky Way revealed by open clusters}",
      journal = {\aap},
         year = 2021,
        month = aug,
       volume = {652},
          eid = {A102},
        pages = {A102},
          doi = {10.1051/0004-6361/202140608},
archivePrefix = {arXiv},
       eprint = {2107.06478},
 primaryClass = {astro-ph.GA},
       adsurl = {https://ui.adsabs.harvard.edu/abs/2021A&A...652A.102H}
}

@ARTICLE{2018RAA....18..146X,
       author = {{Xu}, Ye and {Hou}, Li-Gang and {Wu}, Yuan-Wei},
        title = "{The spiral structure of the Milky Way}",
      journal = {Research in Astronomy and Astrophysics},
         year = 2018,
        month = dec,
       volume = {18},
       number = {12},
          eid = {146},
        pages = {146},
          doi = {10.1088/1674-4527/18/12/146},
archivePrefix = {arXiv},
       eprint = {1810.08819},
 primaryClass = {astro-ph.GA},
       adsurl = {https://ui.adsabs.harvard.edu/abs/2018RAA....18..146X}
}

@ARTICLE{2025AJ....169..115W,
       author = {{Wei}, Xingyin and {Chen}, Jing and {Zhang}, Su and {He}, Feilong and {Zhao}, Yunbo and {He}, Xuran and {Fang}, Yongjie and {Chen}, Xinhao and {Yang}, Hao},
        title = "{Forest Fire Clustering: A Novel Tool for Identifying Star Members of Clusters}",
      journal = {\aj},
         year = 2025,
        month = feb,
       volume = {169},
       number = {2},
          eid = {115},
        pages = {115},
          doi = {10.3847/1538-3881/ada4a0},
       adsurl = {https://ui.adsabs.harvard.edu/abs/2025AJ....169..115W}
}

@ARTICLE{2026ApJ..1001..114W,
       author = {{Wang}, Huajian and {Chen}, Xiaodian and {Wang}, Shu},
        title = "{The Dependence of the Extinction Coefficient on Reddening for Galactic Cepheids}",
      journal = {\apj},
         year = 2026,
        month = apr,
       volume = {1001},
       number = {1},
          eid = {114},
        pages = {114},
          doi = {10.3847/1538-4357/ae552d},
archivePrefix = {arXiv},
       eprint = {2512.23311},
 primaryClass = {astro-ph.GA},
       adsurl = {https://ui.adsabs.harvard.edu/abs/2026ApJ..1001..114W}
}

@ARTICLE{2023MNRAS.526.4107P,
       author = {{Perren}, Gabriel I. and {Pera}, Mar{\'\i}a S. and {Navone}, Hugo D. and {V{\'a}zquez}, Rub{\'e}n A.},
        title = "{The Unified Cluster Catalogue: towards a comprehensive and homogeneous data base of stellar clusters}",
      journal = {\mnras},
         year = 2023,
        month = dec,
       volume = {526},
       number = {3},
        pages = {4107-4119},
          doi = {10.1093/mnras/stad2826},
archivePrefix = {arXiv},
       eprint = {2308.04546},
 primaryClass = {astro-ph.GA},
       adsurl = {https://ui.adsabs.harvard.edu/abs/2023MNRAS.526.4107P}
}

@ARTICLE{2014ApJ...791..131K,
       author = {{Koenig}, X.~P. and {Leisawitz}, D.~T.},
        title = "{A Classification Scheme for Young Stellar Objects Using the Wide-field Infrared Survey Explorer AllWISE Catalog: Revealing Low-density Star Formation in the Outer Galaxy}",
      journal = {\apj},
         year = 2014,
        month = aug,
       volume = {791},
       number = {2},
          eid = {131},
        pages = {131},
          doi = {10.1088/0004-637X/791/2/131},
archivePrefix = {arXiv},
       eprint = {1407.2262},
 primaryClass = {astro-ph.GA},
       adsurl = {https://ui.adsabs.harvard.edu/abs/2014ApJ...791..131K}
}
\bibliographystyle{aasjournalv7}

\end{document}